\documentclass[aps,twocolumn,prd,superscriptaddress,noshowpacs,nofootinbib,noshowkeys,floatfix]{revtex4}
\usepackage[dvips]{graphics,graphicx}
\usepackage{array}
\usepackage{booktabs}
\usepackage[colorlinks=true,linktocpage=true,linkcolor=blue,citecolor=blue]{hyperref}
\usepackage[usenames,dvipsnames]{color}
\usepackage{amsmath, amssymb}
\usepackage{multirow}
\usepackage{longtable}
\usepackage{color}
\usepackage{cleveref}
\usepackage[normalem]{ulem}  % \sout{old text} for strikeout

\renewcommand\sout{\bgroup \color{blue} \ULdepth=-.5ex \ULset}
\newcommand{\sNN}{\sqrt{s_\mathrm{NN}}}
\usepackage{amsmath, amssymb, bm}
\usepackage[sort&compress]{natbib}
\usepackage{graphicx}
\usepackage{lineno}
\usepackage[colorlinks=true]{hyperref}

\begin{document}

\title{{\Large Microscopic study of baryon stopping in low-energy heavy-ion collisions within UrQMD model}}

\author{Sudhir Pandurang Rode}
\email{sudhirrode11@gmail.com, sudhir@jinr.ru}
\affiliation{Veksler and Baldin Laboratory of High Energy Physics, Joint Institute for Nuclear
Research, Dubna, 141980, Moscow region, Russian Federation}

\date{\today}

\begin{abstract}
In low-energy heavy-ion collisions, baryon stopping is an important process, in which protons from the initial colliding nuclei are stopped at the mid-rapidity region. Quantifying such stopping can reveal information on the properties of the nuclear medium, such as net-baryon density. Though experimental measurement of net-proton rapidity spectra provides constraints, microscopic origin of such protons is not fully accessible. Transport model studies can therefore provide deeper insight into the microscopic origin of protons that are transported from initial nuclei, complementing experimental measurements. This article presents an investigation of baryon stopping in minimum-bias Au+Au collisions over a wide range of beam energies, $\sNN = 2.4-17.3$ GeV ($E_{\rm lab} = 1.23A-158A$ GeV). Final-state protons are classified based on their origin by analyzing their interaction history in the UrQMD model. It is found that a significant fraction of transported protons at mid-rapidity originates from initial neutrons rather than initial protons, referred to as \textit{isospin-converted} protons, and their contribution as a function of collision energy and centrality is quantified. Anisotropic flow coefficients of different proton categories are estimated and compared with experimental measurements. Furthermore, a comparison between the $\pi^{-}/\pi^{+}$ ratio and \textit{isospin conversion} rate is performed. The \textit{isospin conversion} rate is further compared with the $\alpha$ parameters of the Kitazawa-Asakawa formalism, revealing a significant deviation from the chemical equilibrium assumption $\alpha_N = \alpha_\pi$ below $\sNN \lesssim$ 10~GeV. Finally, the saturation of the isospin conversion rate is found to coincide with the onset of nuclear transparency, demonstrating that isospin randomization and baryon stopping are coupled phenomena in hadronic transport across the NICA/FAIR energy range.
\end{abstract}

\maketitle

\section{Introduction}
The phase diagram of Quantum Chromodynamics (QCD) remains one of the most fundamental unsolved problems in nuclear physics. Relativistic heavy-ion collisions (HICs) provide a unique opportunity for exploring QCD matter under extreme conditions of temperature ($T$) and baryon chemical potential ($\mu_B$)~\cite{Florkowski:2014yza,UWHeinz,BraunMunzinger:2008tz}. While high energy heavy-ion collisions at RHIC and LHC have established the existence of a deconfined nuclear matter state, quark-gluon plasma (QGP), at high temperatures and low baryon chemical potentials~\cite{rhic1,rhic2,lhc1,lhc2,lhc3,sQGP}, low-energy heavy-ion collisions are essential for mapping the high-$\mu_B$ sector of the QCD phase diagram which is relatively less explored. Heavy-ion collisions at these energies create nuclear matter with baryon densities reaching about 5-10 times the nuclear saturation density. This regime is expected to contain a first-order phase transition between confined hadronic matter and deconfined partonic matter, as well as a critical end point (CEP). The experiments in upcoming accelerator facilities at FAIR~\cite{CBM:2016kpk} and NICA~\cite{nica,MPD:2022qhn,MPD:2025jzd} are expected to probe these regimes. 

A fundamental process governing the initial conditions of these collisions is \textit{baryon stopping}—the process by which baryon number from the colliding nuclei is transported to the mid-rapidity region. Experimentally, baryon stopping is quantified through the net-proton rapidity distribution at mid-rapidity. The degree of stopping determines both the energy deposition into the produced fireball and the net-baryon density, which governs $\mu_B$ and influences possible phase transitions \citep{Mohs:2019xvz,Ivanov:2016xev}. Precise characterization of baryon stopping is, therefore, crucial for interpreting the QCD phase structure searches.

Systematic studies of baryon stopping require measurements across a broad energy range, currently and previously provided by several major experimental programs. The NA49 collaboration at the CERN SPS has delivered comprehensive net-proton rapidity spectra measurements in Pb+Pb collisions at $E_{\rm lab}=$ 20A, 30A, 40A, 80A, and 158A GeV~\cite{NA4920,NA4940,NA49158}. These measurements revealed the transition from single-peaked to double-peaked net-proton rapidity distributions with increasing energy—an indication of changing stopping dynamics. Moreover, the experimental collaborations at AGS also have measured the proton rapidity spectra at energies between $E_{\rm lab}=$2A-8A GeV where the large amount of stopping is expected~\cite{Klay:2003zf,Klay:2001tf}. The E917 collaboration at AGS has calculated the rapidity loss by measuring the net-proton rapidity distributions in $E_{\rm lab}=$6-10.8A GeV~\cite{Back:2000ru}. The experimental measurements suggest that the stopping is substantially less than complete stopping at these energies.  Furthermore, the BRAHMS collaboration quantified the rapidity loss to be, $\delta y = 2.01 \pm 0.14$ using net-proton rapidity distributions at 62.4 GeV, and found out that it exhibits near saturation trend above SPS energies~\cite{BRAHMS:2009wlg}. Looking forward, forthcoming experiments in accelerator facilities under development will probe the highest $\mu_B$ regions in the QCD phase diagram: FAIR (GSI) will study Au+Au collisions at 1A-10A GeV ($\sNN \approx 2-4.5$ GeV) with CBM and HADES experiments, while NICA (JINR) will operate at $\sNN \approx 2.4-11$ GeV with the MPD detector~\cite{nica,MPD:2022qhn,MPD:2025jzd}. The measurement of baryon stopping at these facilities will be essential for mapping the high$-\mu_B$ region of QCD phase diagram.

Many studies have attempted to study baryon stopping by interpreting the experimentally measured net-proton rapidity spectra using various theoretical approaches. The baryon stopping studies using various transport models such as UrQMD model~\cite{bass1998, bleicher1999} and SMASH~\cite{SMASH:2016zqf} have been performed at SPS energies~\cite{Weber:2002qb,Mohs:2019xvz,Mohs:2019iee}. There have been multiple investigations using three-fluid dynamic (3FD) calculations to interpret the net-proton rapidity spectra~\cite{Ivanov:2010cu,Ivanov:2012bh,Ivanov:2013mxa,Ivanov:2015vna,Ivanov:2016xev}. The studies to numerically estimate the stopping fraction using net-proton rapidity spectra have also been performed~\cite{Thakur:2016znw}. The net-baryon rapidity distributions from AGS to LHC energies have been studied within the Non-Uniform Flow Model (NUFM)~\cite{Zhong:2010zz}. Furthermore, a relativistic diffusion approach was used to describe the net-proton rapidity spectra from RHIC to LHC energies in different collision systems~\cite{Kuiper:2006si}.

Experimentally, even though baryon stopping is studied through the measurement of rapidity distribution of net-protons at mid-rapidity, in practice, it is inaccessible to determine whether all net-protons at mid-rapidity originate exclusively from initial protons in the colliding nuclei. In this study, the origin of final-state protons in low-energy heavy-ion collisions is analyzed between $\sNN=$ 2.4-17.3 GeV using the UrQMD model by accessing the interaction history of each final-state particle. This energy span defines the transition from nearly full stopping to partial transparency, covering interesting regime of low-energy heavy ion collisions. It is found that a significant fraction of protons originating from initial nuclei at mid-rapidity originate from neutrons in the initial colliding nuclei rather than initial protons, referred to as \textit{isospin-converted} protons. In heavy-ion collisions, nucleon isospin can be altered relative to their initial state by hadronic interactions, such as charge-exchange reactions and resonance decays. While quantitative effect depends on collision dynamics and is model-dependent, their qualitative presence is well established. The studies by Kitazawa and Asakawa~\cite{Kitazawa:2012at,Kitazawa:2011wh} established that protons experience isospin randomization in the later stages of the collision, where original nucleon isospin distribution becomes blurred due to repeated charge-exchange interactions, such as $p$ + $\pi^{0/-}$ $\longleftrightarrow$ $\Delta^{+/0}$ $\longleftrightarrow$ $n$ + $\pi^{+/0}$. The observations made the authors hold for RHIC energies, above $\sNN\geqslant$ 10 GeV. This article, to the best of the author's knowledge, constitutes one of the first systematic studies to isolate \textit{isospin-converted} protons from the \textit{transported} proton yield across a broad range of energies, $\sNN=$ 2.4-17.3 GeV within the UrQMD model, covering the transition region of the assumptions made by Kitazawa and Asakawa.

The presence of \textit{isospin-converted} protons can affect the interpretation of the results obtained through experimental measurements and implies that conventional net-proton measurements at mid-rapidity include an \textit{isospin-converted} component whose magnitude varies with collision energy and centrality, potentially affecting the interpretation of baryon stopping measurements. Thus, one investigates the effect of isospin conversion as a function of collision energy and centrality. For this study, the version 3.4 of \texttt{UrQMD} model is employed in \texttt{mean-field} (MFD) mode with hard Skyrme equation of state and \texttt{pure transport} (Cascade) mode. The analysis is performed with 20,000 minimum-bias (\texttt{0-14 fm}) Au--Au collisions for each energy between $\sNN=$ 2.4-17.3 GeV. Since \texttt{UrQMD} does not support MFD mode above $\sNN=$ 3.2 GeV ($E_{\rm lab}=$ 4A GeV), only Cascade mode was used after $\sNN=$ 3.2 GeV. Moreover, to validate the predictions, one compares them with available experimental data. 

The paper has a two-part structure. In the first part, a microscopic analysis of baryon stopping and isospin randomization within the UrQMD transport model is presented, introducing a classification of transported protons into those that originate from initial protons (\textit{isospin-preserving} protons) and neutrons (\textit{isospin-converted} protons). One quantifies the isospin conversion rate and its 
dependence on collision energy across $\sNN$ = 2.4--17.3~GeV for Au+Au collisions. In the second part, one studies the sensitivity of experimentally accessible observables, such as rapidity spectra, directed flow slope, and elliptic flow, to this classification. The microscopic isospin conversion rate is also connected to observables measurable at heavy-ion collision experiments. To this end, the isospin conversion rate is compared with predicted and experimentally measured charged pion ratio ($\pi^{-}/\pi^{+}$). Moreover, one compares the $\alpha$ parameters from the Kitazawa-Asakawa (KA) formalism~\cite{Kitazawa:2012at,Kitazawa:2011wh}, which provides isospin corrections to net-proton cumulant measurements, and quantifies the breakdown of the KA chemical equilibrium assumption below $\sNN \lesssim 10$~GeV. Furthermore, one establishes a direct connection between baryon stopping and isospin randomization, demonstrating that the degree of isospin randomization achieved at mid-rapidity is coupled to the onset of nuclear transparency as a function of collision energy, which is relevent for the interpretation of baryon stopping measurements at experiments, such as HADES, BM$@$N, CBM, and MPD.

The paper is organized as follows. In Sec.~II, the UrQMD model is briefly introduced and the approach to classify final state protons into produced and transported categories is outlined. In Sec.~III, the obtained results are presented and discussed, including the energy and centrality dependence of the isospin conversion rate, its comparison with $\alpha$ parameters from the KA formalism, and the coupling between isospin randomization and baryon stopping. Finally, the findings are summarized in Sec.~IV.

\section{Classification in UrQMD model}
The Ultra-relativistic Quantum Molecular Dynamics (UrQMD) model is a microscopic hadronic transport framework designed to simulate the evolution of strongly interacting matter produced in high-energy nucleus-nucleus collisions in the wide range of beam energies~\cite{bass1998,bleicher1999}. Within this framework, interactions at low and intermediate center-of-mass energies per nucleon pair ($\sNN < 5$ GeV) are characterized by stochastic scattering processes involving hadrons and their resonances. At higher energies ($\sNN > 5$ GeV), explicit quark and gluon degrees of freedom are not resolved and particle production is described through excitation of strings and their subsequent fragmentation into hadrons. Nucleus-nucleus collisions within the transport approach are treated as a superposition of all possible binary nucleon-nucleon collisions. Initialization of the target and projectile nuclei involves assigning nucleon positions according to a Woods-Saxon spatial density distribution, for Cascade mode and hard sphere density distribution, for \texttt{MFD} mode. Fermi momenta are assigned stochastically to each nucleon within the rest frame of its respective nucleus. A collision between two particles occurs if their relative transverse distance, $d_{\text{trans}}$ falls below a critical value determined by the total interaction cross-section ($\leqslant d_{0} = \sqrt{\frac{\sigma_{tot}}{\pi}}$). The model includes experimental information such as hadronic cross-sections, resonance decay widths and decay modes. By default, UrQMD operates in Cascade mode, where propagation of particles between subsequent collisions is carried out in straight line trajectories with their velocities. In addition to Cascade mode, UrQMD allows nuclear MFD potential based on the hard Skyrme equation of state. The choice of MFD is limited to incident beam-energies below 4.0 GeV/nucleon, where nuclear density is the highest and nuclear MFD plays a significant role in the collective dynamics of the medium.

In the UrQMD model, the final-state and spectator particle information is stored in the output files \texttt{f13} and \texttt{f14}. While \texttt{f14} contains the four-momenta, particle identities and properties, such as mass, \texttt{f13}, additionally, includes phase space information at the freeze-out of all particles. The number of collisions, $N_{\rm coll}$ in the files represents the number of interactions each particle underwent during evolution of the system, with spectator particles having, $N_{\rm coll}$ $=$ 0. The complete interaction history of each particle, including intermediate interactions is recorded in \texttt{f15}. In addition, the \texttt{f15} file contains comprehensive information regarding all particle creation and annihilation events, including unique identifiers for each particle and its parent(s), as well as status flags indicating the nature of each interaction. Hence, to identify whether a final-state proton is produced or transported from initial nucleons, one combines the information from both \texttt{f13} and \texttt{f15} outputs. The UrQMD model is employed in its standard form without any modifications to the underlying source code. All analysis is performed externally on the model output files. One begins by extracting the phase space information of the final-state proton from \texttt{f13} output file. The complete interaction history of each particle is then accessed by iterating through the collision record, using $N_{\rm coll}$ and parent particle identifiers to locate each preceding interaction. At each step, only non-strange baryonic intermediate states (e.g., $\Delta$, $N^*$) corresponding to particle type indices, $i_{typ} < 26$ in the UrQMD particle table are considered, ensuring the interaction history does not possess any strange baryonic states. Therefore, the final-state proton is considered to be transported if all ancestors in its interaction history are non-strange baryons. Accordingly, the final-state proton is classified as produced if any strange baryon or meson appears in its ancestral history. The types of the interactions, denoted by their respective processes IDs are described in the UrQMD manual~\cite{manual}. 

The protons that are transported can originate from both initial proton and initial neutron. In the former case, the isospin remains unchanged, whereas, in the latter case, the initial neutron has undergone an isospin change to appear as a final-state proton. This allows the transported protons to be further categorized on the basis of isospin conversion history, which is performed by tracking each interaction in the collision history of the final-state proton.

\section{Results and Interpretation}

The analysis of the interaction history of final-state protons reveals that a significant fraction of protons originating from initial nucleons at mid-rapidity in Au-Au collisions originates from initial neutrons rather than initial protons. This motivates the classification of final-state protons into the following categories based on their origin: 
\begin{itemize}
\item If the final-state proton originates from initial nucleon, it is termed as \textit{transported} proton.
\item If it specifically originates from initial proton, it is termed as \textit{isospin-preserving} proton. 
\item However, if it originates from initial neutron, it is referred to as \textit{isospin-converted} proton. 
\item Furthermore, the one produced during the collision evolution is termed as \textit{produced} proton. 
\end{itemize}
Thus, \textit{transported} protons are composed of both \textit{isospin-preserving} and \textit{isospin-converted} protons.

\begin{figure*}[!t]
\includegraphics[scale=0.38]{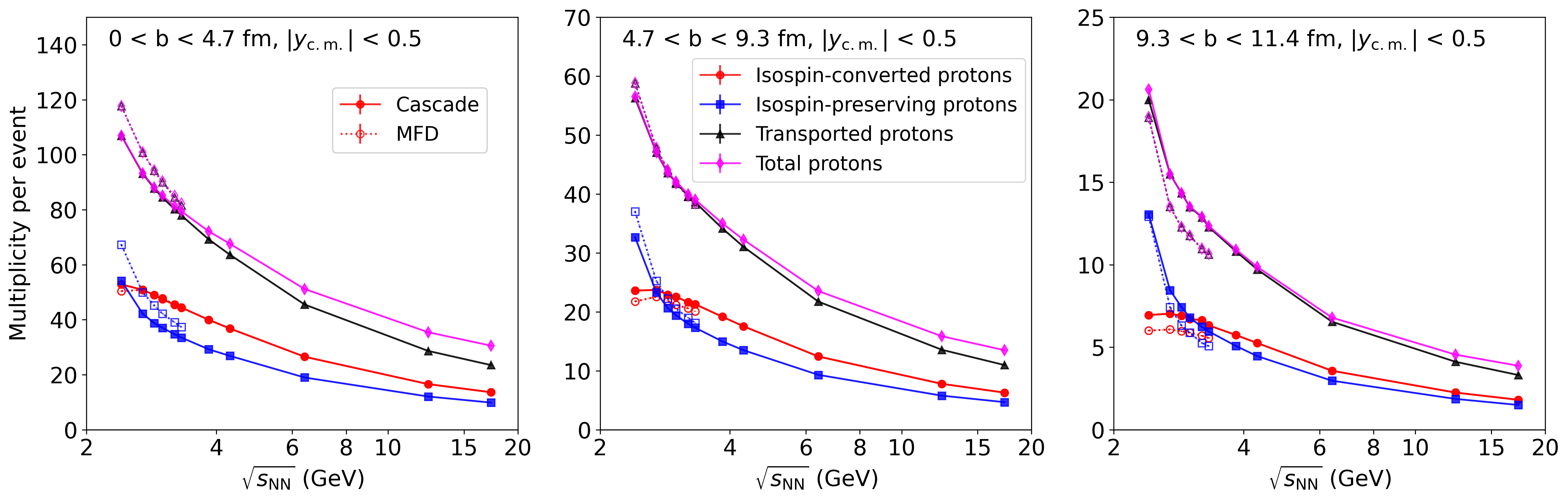}
\includegraphics[scale=0.38]{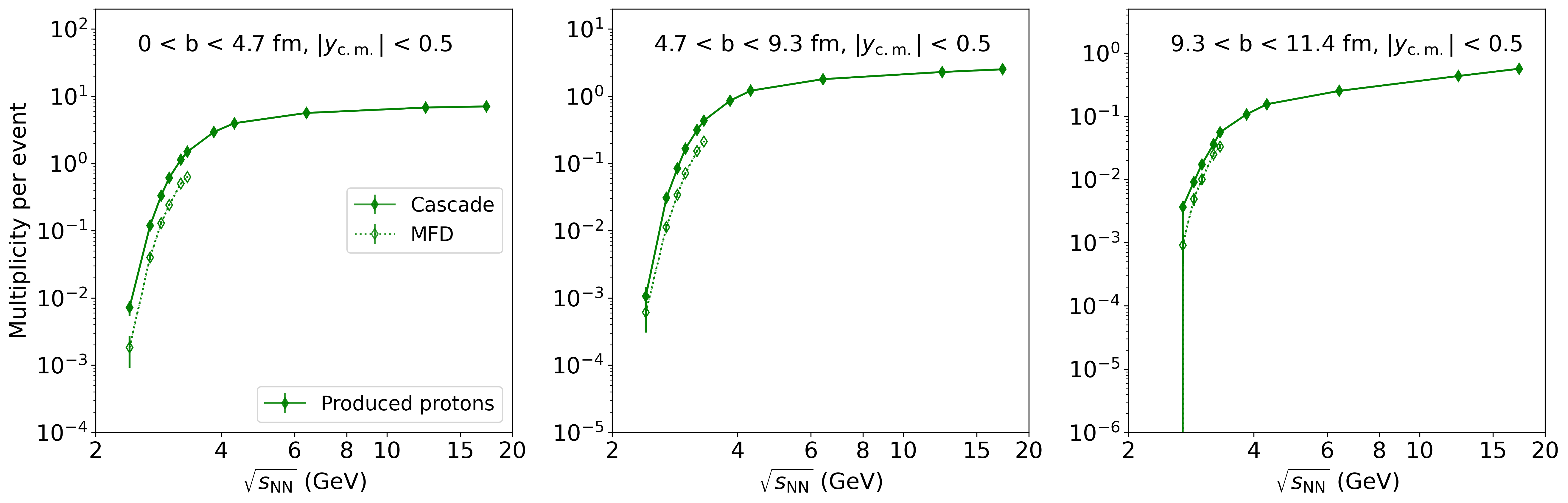}
\caption{Comparison of multiplicities of differently categorized protons per event as a function of $\sNN$ in three impact parameter intervals (0 $<b<$ 4.7 fm, 4.7 $<b<$ 9.3 fm and 9.3 $<b<$ 11.4 fm) using UrQMD model with MFD (markers with dotted lines) and Cascade mode (markers with solid lines). Upper panel shows \textit{transported} proton multiplicities and lower panel shows \textit{produced} proton multiplicities with increasing impact parameter intervals from left to right plots.}
\label{f1}
\end{figure*}

Initially, the multiplicities of all categorized protons are evaluated and compared as a function of collision energies in $|y_{\rm c.m.}|<$ 0.5 across three different impact parameter intervals, 0 $<b<$ 4.7 fm, 4.7 $<b<$ 9.3 fm and 9.3 $<b<$ 11.4 fm, as shown in Figure~\ref{f1}. These impact parameter regions approximately belong to centrality classes, 0-10$\%$, 10-40$\%$ and 40-60$\%$ at HADES energy, $\sNN =$ 2.4 GeV and one uses same intervals at all studied energies~\cite{HADES:2017def}. The upper panel of Figure~\ref{f1} shows the multiplicities of all categorized protons, except for \textit{produced} protons which is shown in the lower panel. From the upper panel, it can be observed that yields of all categorized protons decrease as beam energy increases across impact parameter intervals with steady fall at low energies in the case of \textit{isospin-converted} protons and sharp fall in the case of rest of the categorized protons. The yields of all categorized protons are sensitive to the mode of UrQMD, MFD and Cascade. The magnitudes of the yield decrease faster in MFD than Cascade as impact parameter increases, transitioning from higher to lower values with respect to Cascade mode for all protons, except \textit{isospin-converted} protons. On the other hand, as shown in the lower panel, the \textit{produced} protons multiplicities increase monotonously as a function of collision energy with negligible multiplicity at the lowest energy in the analysis. The yield in the central collisions is the highest and decreases as the impact parameter increases. It is confirmed that no anti-protons are produced below $\sNN =$ 3.84 GeV within statistical precision, ruling out pair production as a mode of production for \textit{produced} protons. Strange baryons seem to be the main source of production. The yield of \textit{produced} protons using MFD is always smaller than Cascade mode.

\begin{figure*}[!t]
\includegraphics[scale=0.38]{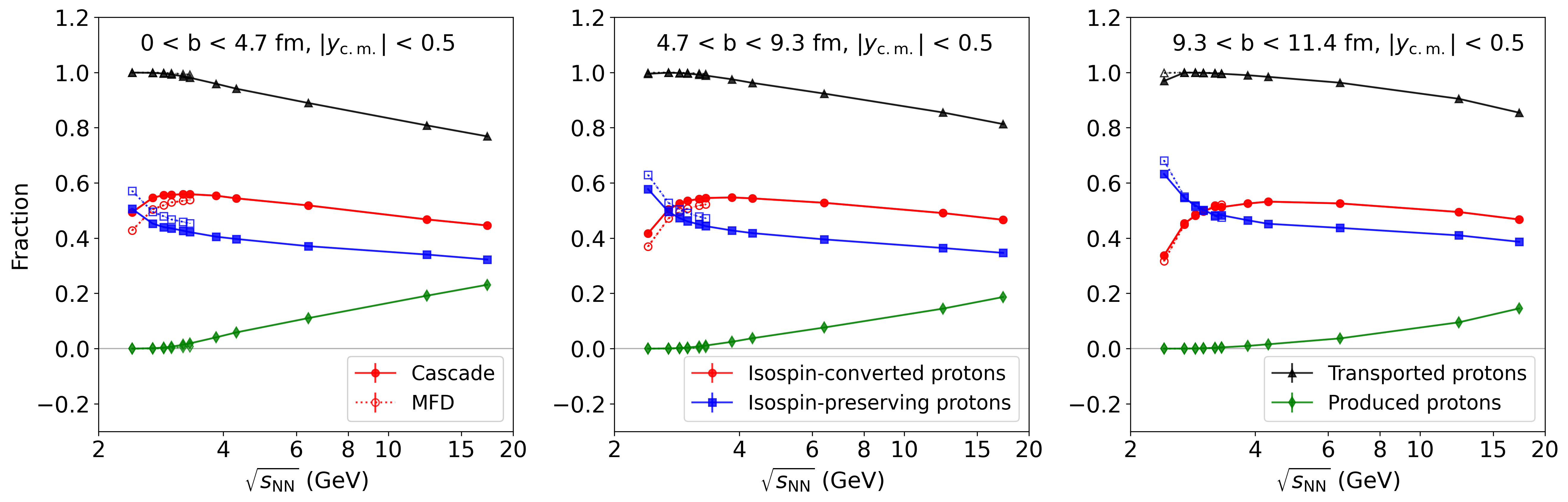}
\caption{Comparison of fractions of differently categorized protons with respect to total final-state protons as a function of $\sNN$ in three impact parameter intervals (0 $<b<$ 4.7 fm, 4.7 $<b<$ 9.3 fm and 9.3 $<b<$ 11.4 fm) using UrQMD model with MFD (markers with dotted lines) and Cascade mode (markers with solid lines).}
\label{f2}
\end{figure*}

\begin{figure*}
\includegraphics[scale=0.38]{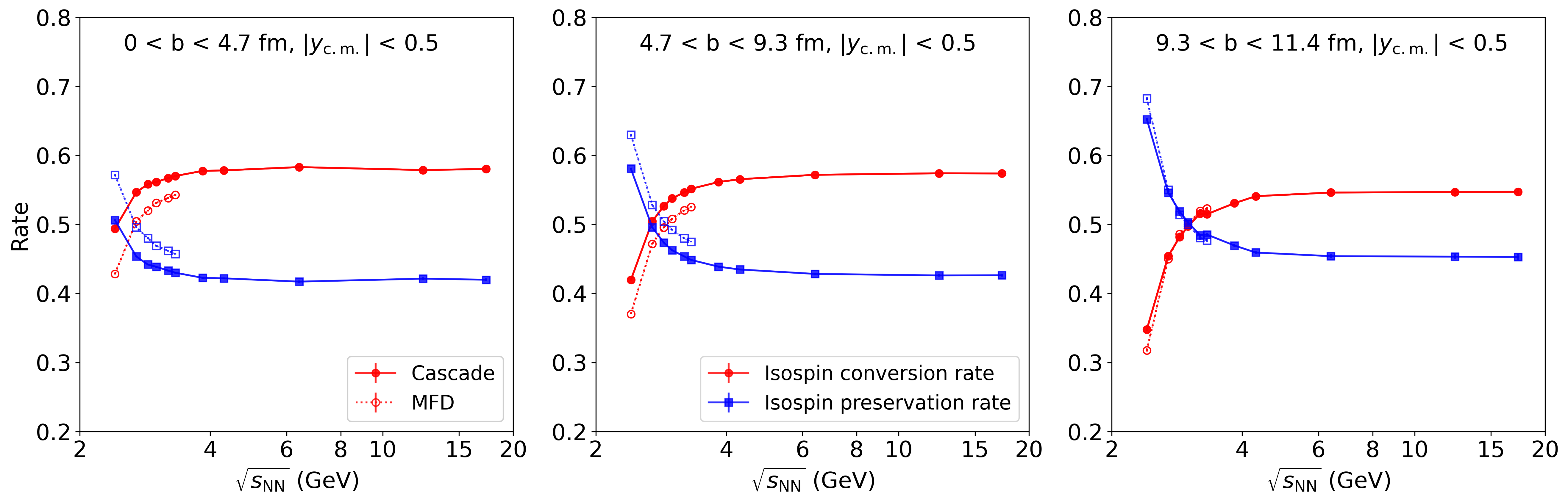}
\caption{Comparison of \textit{isospin conversion} and \textit{isospin preservation} rate as a function of $\sNN$ in three impact parameter intervals (0 $<b<$ 4.7 fm, 4.7 $<b<$ 9.3 fm and 9.3 $<b<$ 11.4 fm) using UrQMD model with MFD (markers with dotted lines) and Cascade mode (markers with solid lines).}
\label{f3}
\end{figure*}

Next, one determines the fraction of \textit{transported} protons, \textit{isospin-preserving} protons, \textit{isospin-converted} protons and \textit{produced} protons with respect to total final-state protons. These fractions were estimated in $|y_{\rm c.m.}|<$ 0.5 as a function of collision energies across same three different impact parameter intervals as shown in Figure~\ref{f2}. Both \textit{transported} protons and \textit{produced} protons fractions vary monotonically with collision energies across impact parameter intervals, with the former decreasing and the latter increasing with increasing collision energy, $\sNN$. Moreover, both fractions vary the most in central collisions, in comparison to peripheral collisions. The \textit{isospin-preserving} proton and \textit{isospin-converted} proton fractions show sensitivity to the MFD and Cascade modes, with the former decreasing monotonically as a function of $\sNN$ in the both modes. The latter behaves non-monotonically with initial short rise at low energies in both Cascade and MFD modes, with gradual decreasing trend as $\sNN$ increases above $\approx$3-4 GeV in case of Cascade mode. The \textit{isospin-preserving} proton fractions shows more yield in MFD mode as opposed to Cascade mode, whereas, \textit{isospin-converted} proton fraction exhibit the opposing nature. The disagreement of \textit{isospin-preserving} and \textit{isospin-converted} proton fractions between Cascade and MFD mode decrease as impact parameter increase, with full agreement in the peripheral collisions. The \textit{isospin-converted} proton fraction lies between 0.4-0.6 at all studied energies, suggesting strong isospin-exchange. Note that the energy dependence in \textit{isospin-preserving} and \textit{isospin-converted} proton fractions above $\approx$3-4 GeV is due to increasing contribution of \textit{produced} protons in total final-state protons. However, the relative fractions of these protons with respect to \textit{transported} proton yield, referred as, \textit{isospin conversion} and \textit{isospin preservation} rates, begin to saturate above $\sNN\approx$ 4 GeV, as shown in Figure~\ref{f3}. Above this energy, the \textit{isospin conversion} rate saturates at lower values as impact parameter increases, suggesting effect of finite hadronic phase lifetime as well as limited charge exchange, leading to incomplete isospin randomization. The rate saturates around $\approx$~0.58 in central collisions, which is close to neutron to nucleon ratio in Au nucleus (N/A $\approx$ 0.6), hinting that the system is close to complete isospin randomization. Small rate below $\sNN\approx$ 4 GeV suggests the limiting pion density, leading to reduced \textit{isospin conversion} rate. Moreover, as energy increases, increase in symmetric pion production leads to more charge exchange processes, enhancing \textit{isospin conversion} rate.

\begin{figure*}
\begin{center}
\includegraphics[scale=0.38]{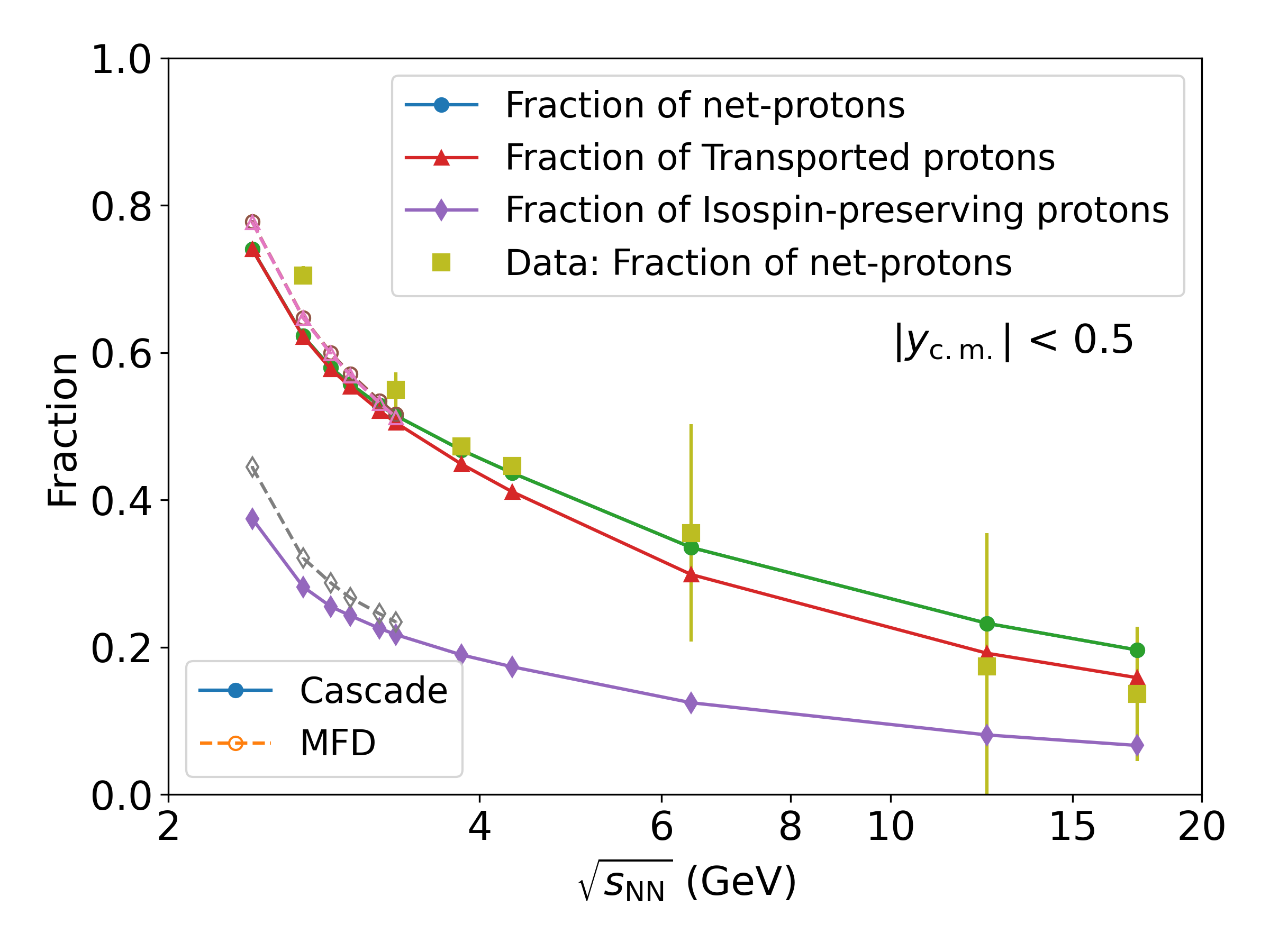}
\includegraphics[scale=0.38]{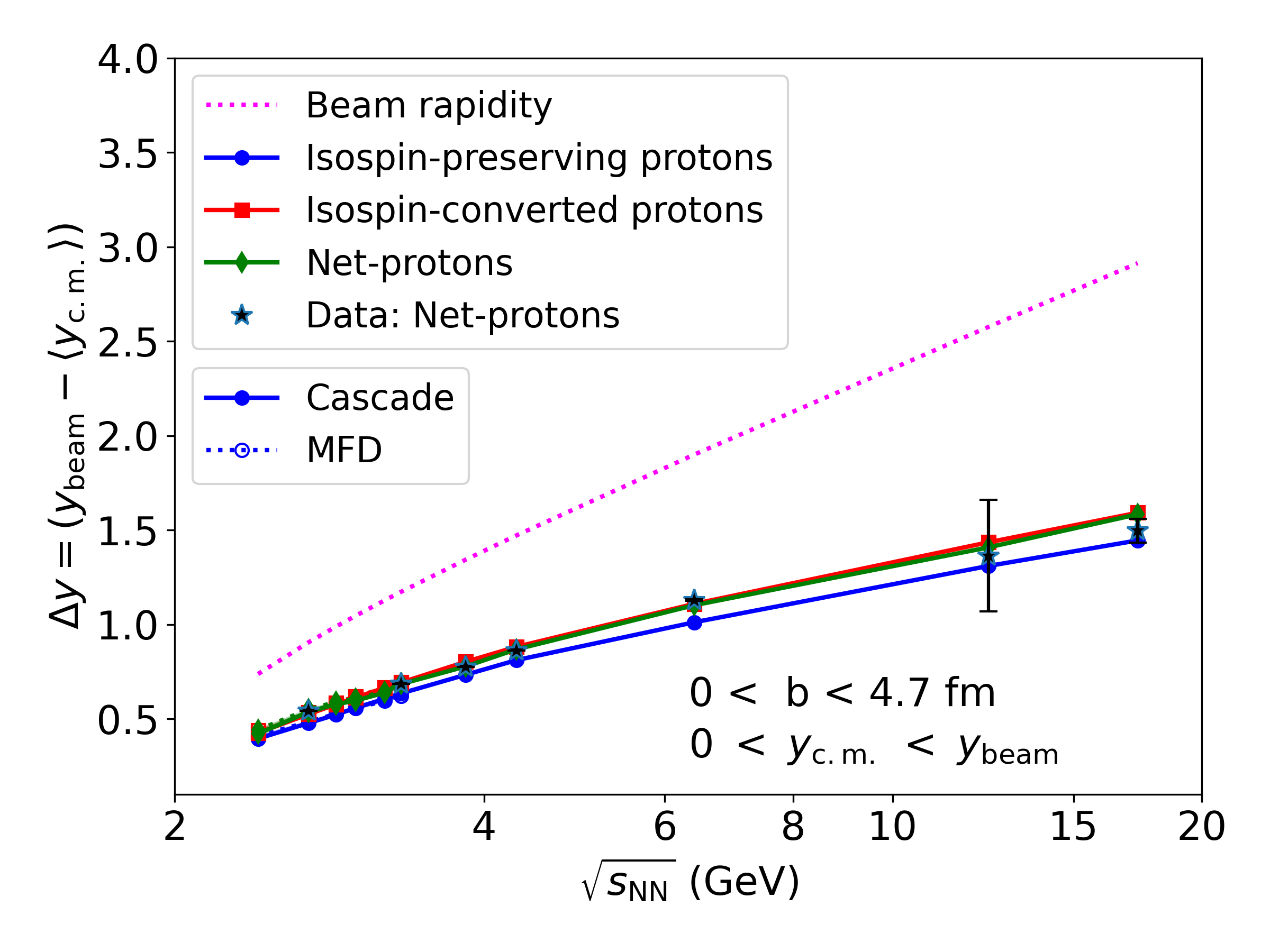}
\includegraphics[scale=0.38]{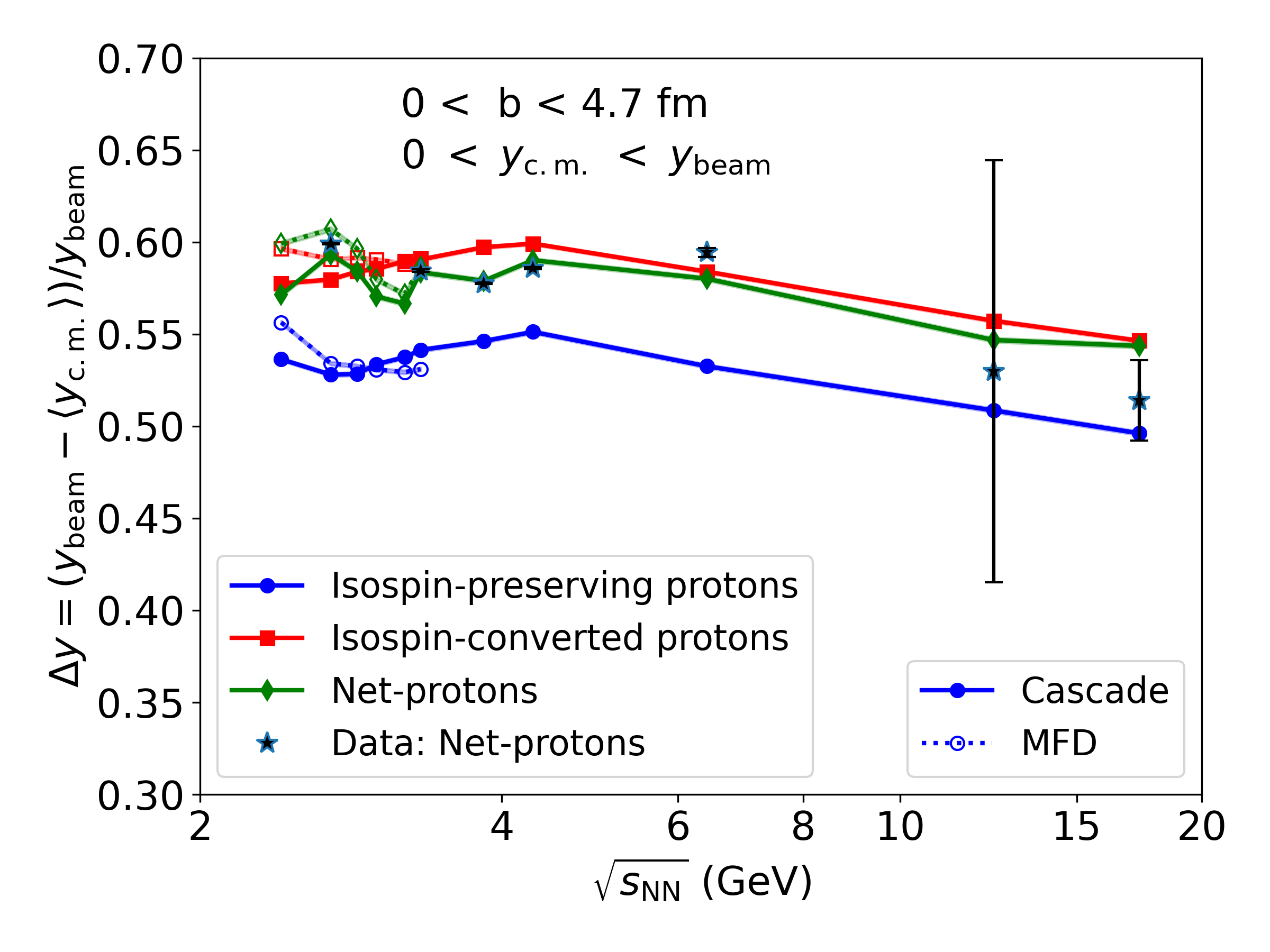}
\caption{Comparison of various fractions of stopped/transported protons in mid-rapidity (top left) and absolute (top right) and relative (bottom) rapidity loss of net-protons and \textit{transported} protons as a function of $\sNN$ in the central Au-Au collisions using UrQMD model with MFD (markers with dotted lines) and Cascade mode (markers with solid lines). The fraction of net-protons and rapidity loss are calculated for experimental data and compared with the predictions.}
\label{f4}
\end{center}
\end{figure*}

One now moves on to estimate the fraction of net-protons, estimated as the ratio of number of net-protons in $|y_{\rm c.m.}| <$ 0.5 to those within $|y|<$ $y_{\rm beam}$. This fraction is being compared with the fraction computed from experimentally measured rapidity spectra of net-protons. Moreover, one also estimates the fraction of \textit{transported} protons and \textit{isospin-preserving} protons: the ratio of number of such protons in $|y_{\rm c.m.}| <$ 0.5 to the number of net-protons within $|y_{\rm c.m.}|<$ $y_{\rm beam}$. The comparison is shown in the left plot of Figure~\ref{f4}. The fraction of net-protons from both model and experimental measurements agree with each other within uncertainties. The fraction of \textit{transported protons} also agrees with fraction of net-protons reasonably well, which confirms the net-protons to be a good proxy to estimate the \textit{transported protons}. As a reference, one also estimates the fraction of \textit{isospin-preserving} protons for comparison and it is seen that the fraction is about 2 times smaller than net-protons and that the difference decreases with an increase in energy with an eventual agreement with the data within uncertainties for the last two energies. Furthermore, the average rapidity loss, $\langle\delta y\rangle$ $=$ $y_{\rm beam}$ - $\langle y_{\rm c.m.}\rangle$, of net-protons, \textit{isospin-preserving} protons and \textit{isospin-converted} protons in UrQMD are also evaluated. The mean $\langle y_{\rm c.m.}\rangle$ is estimated within 0 $<$ $y_{\rm c.m.}$ $<$ $y_{\rm beam}$ region. In case of measured rapidity loss, $\langle y_{\rm c.m.}\rangle$ is obtained after fitting the net-proton spectra using two source fit within 0 $<$ $y_{\rm c.m.}$ $<$ $y_{\rm beam}$ region. It is found that \textit{isospin-converted} protons undergo more rapidity loss compared to \textit{isospin-preserving} protons. The rapidity loss measured by E917 collaboration~\cite{Back:2000ru} has different kinematic cuts than what one has used and hence, one does not compare quantitatively. However, the average rapidity loss of measured net-protons at AGS and SPS energies are estimated, except for $E_{\rm lab}=$ 10.8A GeV. It is found that the average rapidity loss varies linearly with beam energy. The predictions for net-protons and \textit{isospin-converted} protons show agreement with each other and the measured loss. The loss tends to deviate from beam rapidity as beam energy increases. The comparison is shown in the right plot of the Figure~\ref{f4}. One also evaluates relative rapidity loss, defined with respect to beam rapidity, which is also measured by E917 collaboration and similar to their observation an energy independent behaviour is observed in the similar energy interval, as well as a hint of energy dependence at high energies, as shown in the bottom plot of the Figure~\ref{f4}.

\begin{figure*}[!t]
\includegraphics[scale=0.38]{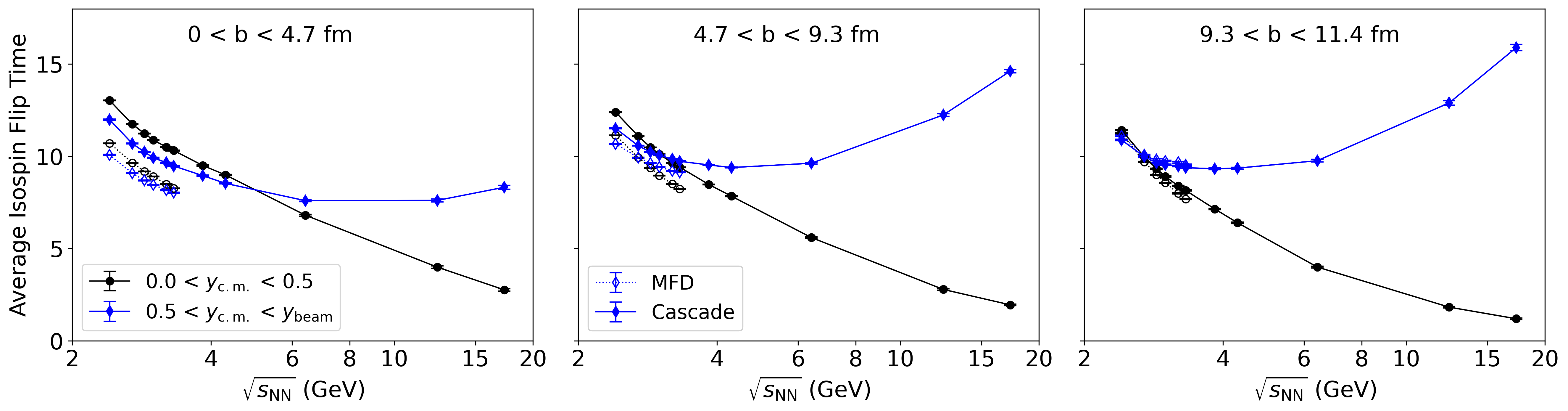}
\includegraphics[scale=0.38]{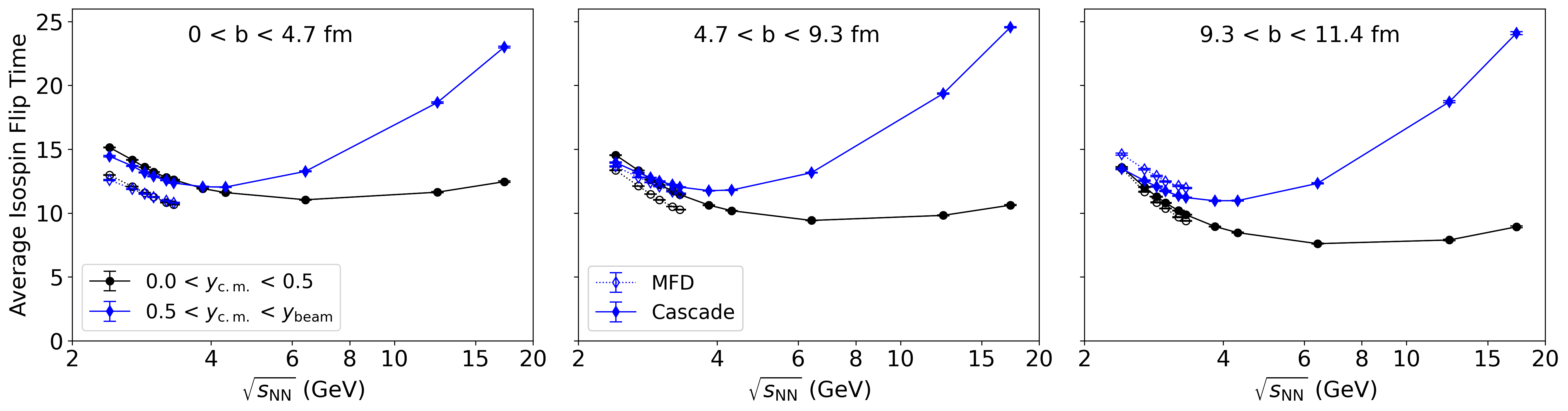}
\caption{Average isospin \textit{flip time} of \textit{isospin-converted} protons as a function of $\sNN$ for different impact parameter regions (0 $<b<$ 4.7 fm, 4.7 $<b<$ 9.3 fm and 9.3 $<b<$ 11.4 fm). The top panel is for direct conversion and bottom panel is for indirect conversion via intermediate resonance.}
\label{f5}
\end{figure*}

\begin{figure*}
\includegraphics[scale=0.38]{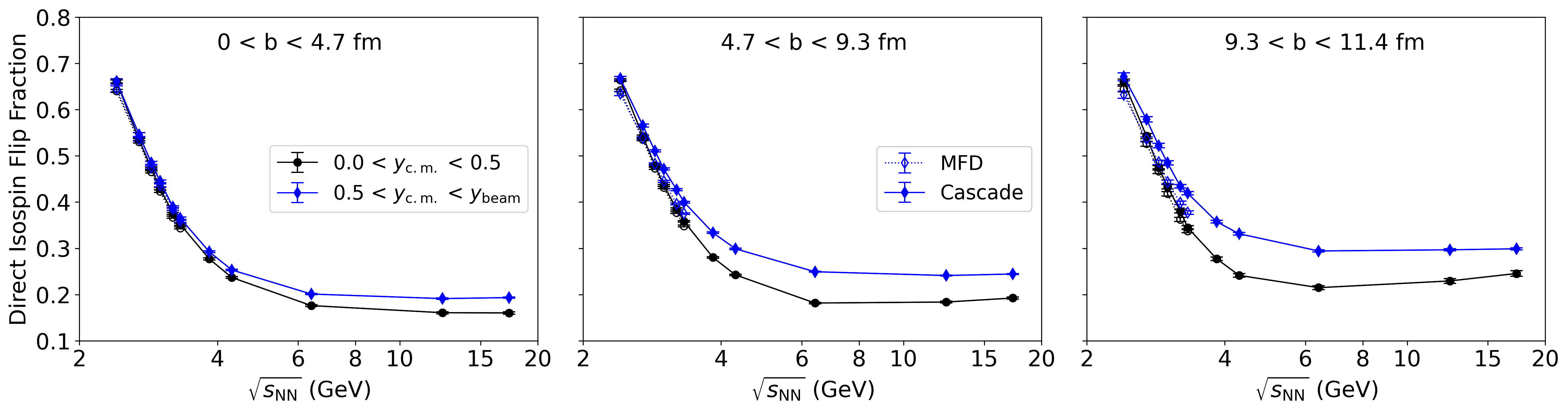}
\caption{Fraction of direct isospin conversions with respect to total \textit{isospin-converted} protons as a function of energy for central (b = 0-4.7 fm), mid-central (b = 4.7-9.3 fm), and peripheral (b = 9.3-11.4 fm) Au+Au collisions.}
\label{f5_1}
\end{figure*}

The observed existence of non-negligible amount of \textit{isospin-converted} protons suggests that such protons have undergone isospin conversion. To understand this further, one estimates average \textit{flip time}, the instance in the evolution when the \textit{isospin-converted} proton appears. Note that isospin conversion can occur on multiple occasions, but only the last instance for a proton is recorded. Moreover, it can occur either directly through single interaction, such as, $N\Delta \rightarrow NN$ and $NN \rightarrow N\Delta$ or indirectly via any intermediate $\Delta$ resonance. Therefore, one estimates the average \textit{flip time} for both such scenarios as a function of $\sNN$ in $|y_{\rm c.m.}| <$ 0.5 and 0.5 $<|y_{\rm c.m.}| <$ $y_{\rm beam}$, for three impact parameter regions (0 $<b<$ 4.7 fm, 4.7 $<b<$ 9.3 fm and 9.3 $<b<$ 11.4 fm), as shown in Figure~\ref{f5}.  The upper panels show the average \textit{flip time} of isospin conversion through direct interaction, whereas, lower panels are for indirect conversion. The average \textit{flip time} for direct conversions at mid-rapidity, $|y_{\rm c.m.}| <$ 0.5 shows monotonic behaviour and decreases with increasing energy for all three impact parameter intervals. However, in the forward rapidity region, the monotonic behaviour is not observed in any impact parameter interval.  The average \textit{flip time} show a non-monotonic trend where it decreases initially, with eventual increasing behaviour as energy increases. These trends in both rapidity regions create an interesting diversion around $\sNN\approx$~3-4 GeV energy range depending on the impact parameter interval. The absolute values of \textit{flip time} lie between 1-14 fm and 8-16 fm for mid-rapidity and forward region, respectively. In comparison, average \textit{flip time} for indirect conversions is larger and it lies between 10-15 fm for mid-rapidity region, and 10-25 fm for forward region, reflecting longer reaction duration due to multiple intermediate interactions involving resonances in indirect conversions. In both rapidity and all centrality regions, the trend is non-monotonic, with similar diversion as seen in case of direct conversion between rapidity regions around $\sNN\approx$~3-4 GeV. Given the interesting sensitivity of the average \textit{flip time} to the type of interactions leading to isospin conversion, the fraction of \textit{isospin-converted} protons produced through direct interaction with respect to the total number of \textit{isospin-converted} protons is estimated, as a function of energy for both rapidity regions in three impact parameter regions, as shown in Figure~\ref{f5_1}. About 65$\%$ of the \textit{isospin-converted} protons originate through direct conversion at the lowest energy, $\sNN=$ 2.4 GeV in all impact parameter regions, hinting towards limited pion density. This fraction decreases sharply after that to $\approx$~25-30$\%$ as beam energy reach around $\sNN\approx$ 4 GeV, exhibiting strongly changing collision dynamics. The fraction further saturates above this energy, with indirect conversions taking over at higher energies where symmetric pion production increases. The fraction estimated in two rapidity regions shows diversion above energy region between $\sNN\approx$~3-4 GeV with a degree of disagreement increasing as impact parameter increases.

\begin{figure*}
\begin{center}
\includegraphics[scale=0.45]{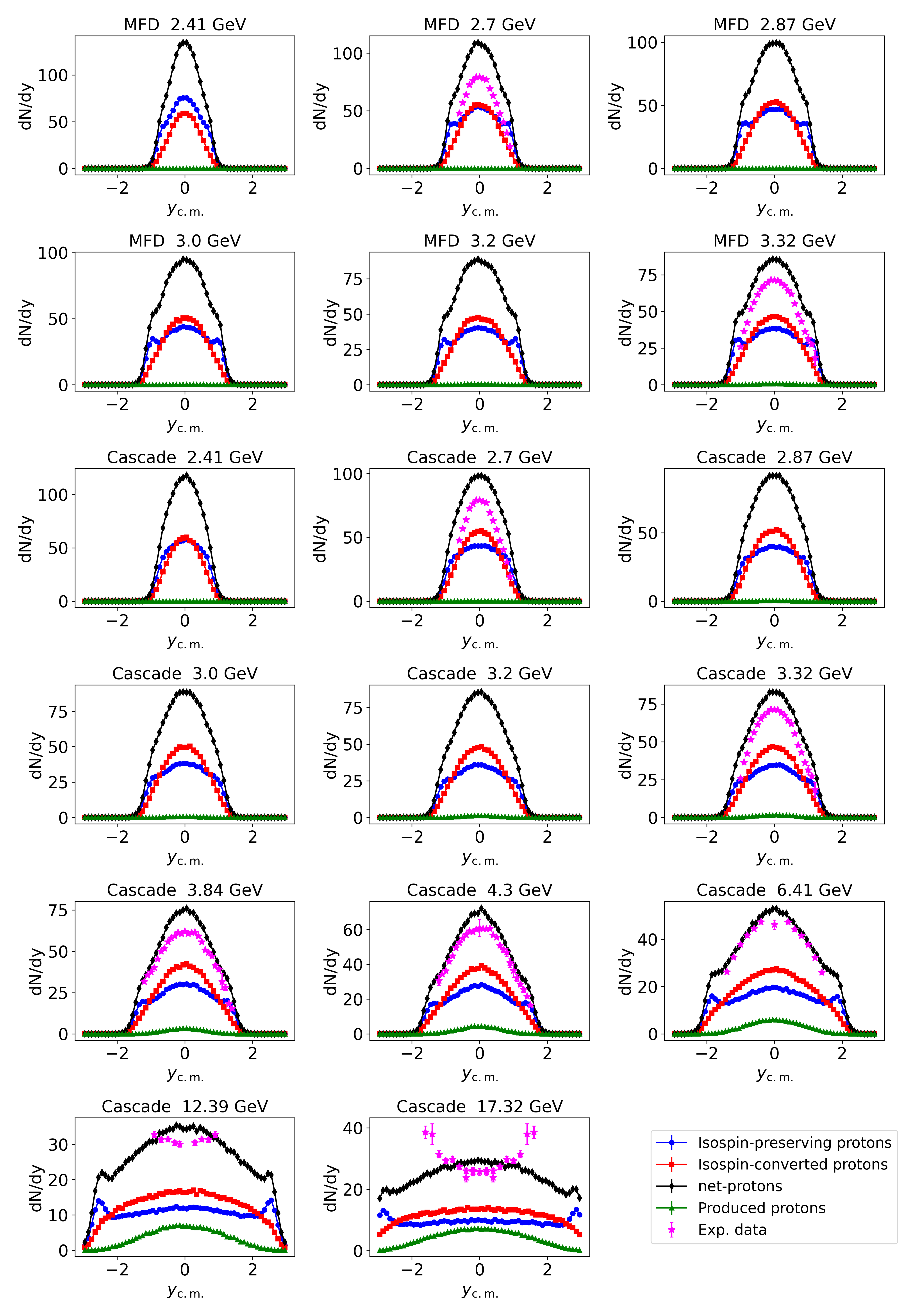}
\caption{UrQMD predictions for rapidity distributions of \textit{isospin-preserving}, \textit{isospin-converted}, \textit{produced} protons and net-protons at different $\sNN$ in central (0 $<b<$ 4.7 fm) Au-Au collisions for both MFD and Cascade mode and their comparison with available experimental measurements.}
\label{f6}
\end{center}
\end{figure*}

\begin{figure}
\includegraphics[scale=0.5]{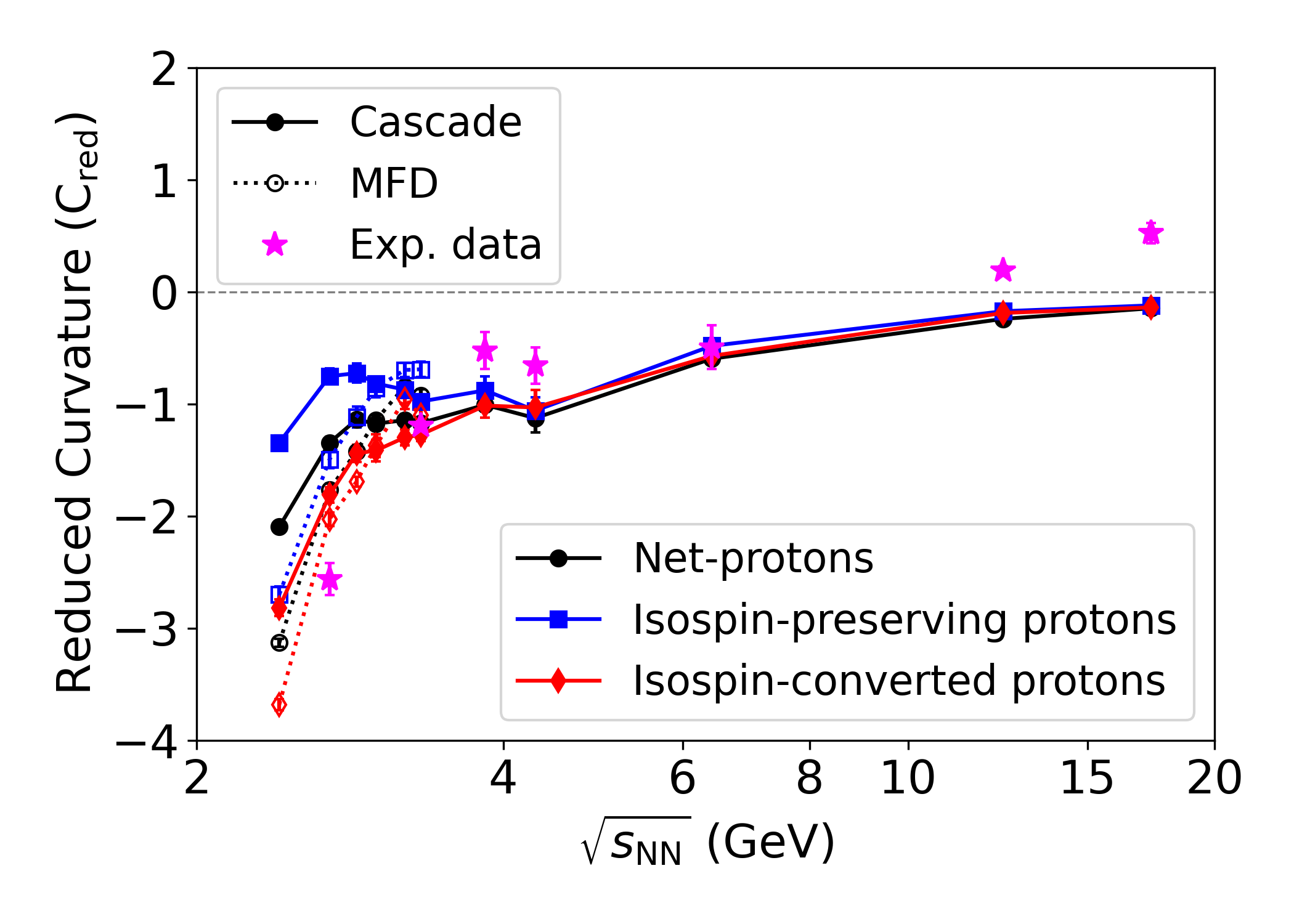}
\caption{UrQMD predictions of reduced curvature of rapidity spectra at mid-rapidity of \textit{isospin-preserving}, \textit{isospin-converted} protons and net-protons as a function of $\sNN$ in central (0 $<b<$ 4.7 fm) Au-Au collisions in both MFD and Cascade mode.}
\label{f7}
\end{figure}

As mentioned earlier, experimentally, the study of baryon stopping is accessible through net-proton rapidity distributions and therefore, the corresponding spectra of net-protons, \textit{isospin-converted} protons, \textit{isospin-preserving} protons and \textit{produced} protons are computed within UrQMD model in central (0 $<$ b $<$ 4.7 fm) Au-Au collisions. The spectators are excluded while estimating this observable by requiring $N_{\rm coll}$ $>$ 0. The comparison of distributions for different energies for both MFD and Cascade modes are shown in Figure~\ref{f6}. The prediction were compared with available measured net-proton rapidity spectra at AGS ($\sNN=$ 2.7, 3.32, 3.84, 4.3 GeV) and SPS ($\sNN=$ 6.41, 12.39, 17.32 GeV) energies. The predictions overestimate the measured dN/dy at all energies, with contrasting mid-rapidity shapes at $\sNN=$ 12.39 and 17.3 GeV.  One notes that the peaks near the beam rapidity does not appear in the case of \textit{isospin-converted} protons, potentially, suggesting more rapidity loss in comparison to \textit{isospin-preserving} protons. Moreover, the results exhibit more \textit{isospin-preserving} proton yield in comparison to \textit{isospin-converted} protons in the forward rapidity region. To quantify the nature of rapidity distributions at mid-rapidity, one evaluates the double derivative, i.e., global maxima or minima, referred to here as reduced curvature ($\rm C_{red}$)~\cite{Ivanov:2010cu} of the rapidity spectra of net-protons, \textit{isospin-converted} protons, \textit{isospin-preserving} protons for both predictions and experimental measurements. The calculations are performed within $|y_{\rm c.m.}|<$ 0.5 (0.8 at $\sNN=$ 6.41 GeV and 1.0 at $\sNN=$ 12.39 and 17.3 GeV), using polynomial of order 3. The comparisons are shown in Figure~\ref{f7} as a function of $\sNN$. The reduced curvatures corresponding to predicted rapidity spectra of differently categorized protons show sensitivity to the type of proton as well as nuclear potential. MFD mode predicts smaller values of $\rm C_{red}$ compared to Cascade mode at the lowest energy, with decrease in the difference as energy increases. The observable, $\rm C_{red}$ for net-protons lie between \textit{isospin-preserving} and \textit{isospin-converted} protons, forming an origin-based ordering. Eventually, $\rm C_{red}$ becomes insensitive to the proton type above $\sNN\approx$ 4 GeV. $\rm C_{red}$ of measured rapidity spectra show similar trend as predictions, with UrQMD results at $\sNN=$ 3.32 and 6.41 GeV agreeing with the measured $\rm C_{red}$. The results at $\sNN=$ 3.84 and 4.3 GeV slightly underestimate the experimental data.  Above $\sNN=$ 6.41 GeV, the UrQMD fails to reproduce the mid-rapidity shape. The reduced curvature for \textit{isospin-preserving} protons show a non-monotonic behaviour as a function energy with a slight saturation around $\sNN\approx$~3-4 GeV. Similar trend is seen for $\rm C_{red}$ of measured spectra, with hint of saturation around $\sNN\approx$~4-6 GeV. On the other hand, for net-protons and \textit{isospin-converted} protons, the trend is reasonably monotonic, with complete agreement between all three categories beyond $\sNN=$4 GeV.

\begin{figure*}
\includegraphics[scale=0.38]{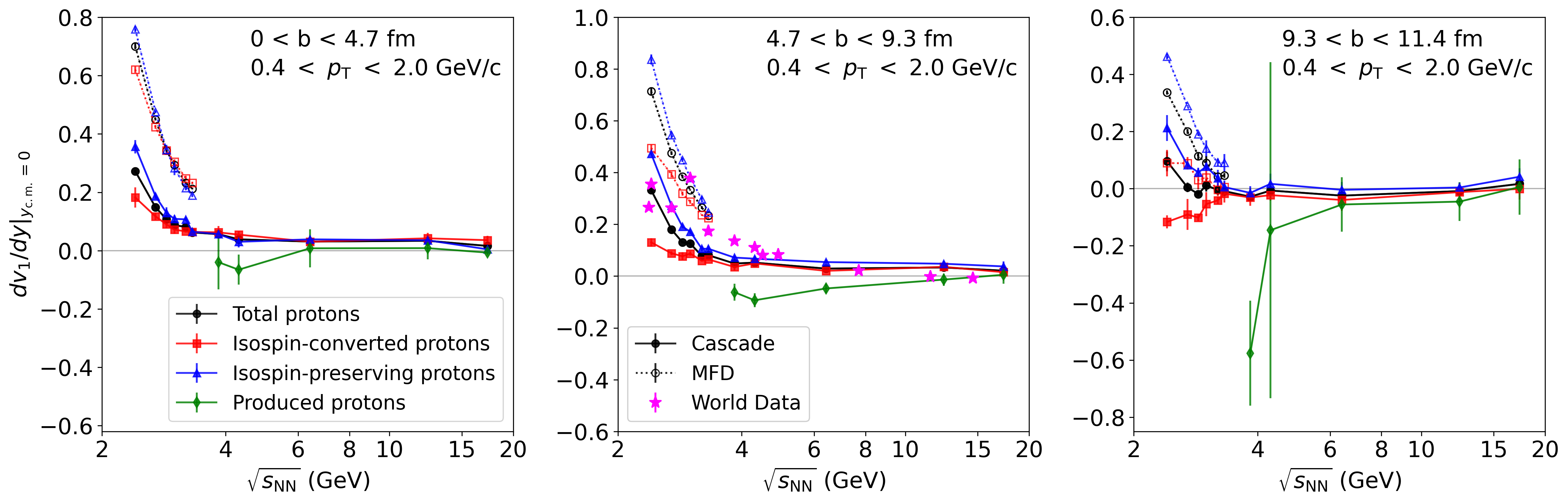}
\caption{Comparison of UrQMD prediction of slope of directed flow of \textit{isospin-preserving}, \textit{isospin-converted}, \textit{transported} and \textit{produced} protons as a function of $\sNN$ in different impact parameter collisions in both MFD and Cascade mode. The predictions are compared with experimental measurements in non-central collisions (middle plot).}
\label{f10}
\end{figure*}

Moving on, \textit{isospin-converted} protons, \textit{isospin-preserving} protons and \textit{produced} protons, originate from different stages of the collision and have different interaction histories. Since \textit{transported} protons originate from initial nuclei, they are expected to carry memory of early collision geometry, whereas, \textit{produced} protons reveal the dynamics of the evolution of the dense medium. Thus, to characterize the collective behaviour of these categorized protons, it is interesting to investigate their various anisotropic flow coefficients as a function of energy. Initially, the slope of directed flow of all categorized protons is evaluated as a function of collision energies as shown in Figure~\ref{f10}, in different impact parameter regions (0 $<b<$ 4.7 fm, 4.7 $<b<$ 9.3 fm, and 9.3 $<b<$ 11.4 fm) approximately belonging to centrality classes, 0-10$\%$, 10-40$\%$ and 40-60$\%$~\cite{HADES:2017def}. The slope of the predicted directed flow is estimated by fitting $v_{1}(y)$ at the mid-rapidity ($y_{\rm c.m.}<0.8$) in 0.4 $<$ $p_{\rm T}$ $<$ 2.0 GeV/c using polynomial, $\rm a_1y + a_3 y^3$. The predictions for mid-central collisions are compared with experimental measurements~\cite{FOPI:2004bfz,E895:2000maf,HADES:2020lob,STAR:2020dav,STAR:2014clz,E895:1999ldn}, as shown in middle plot. The predictions from Cascade mostly underestimate experimental measurements at intermediate energies, with a reasonable agreement at the lowest and the highest energies, except for 3 GeV, where MFD prediction agrees with the data. The predictions from MFD are always higher than Cascade ones across energies and impact parameters. Similar to reduced curvature, the predictions show an origin-based ordering in the slope of directed flow at all centralities decreasing from \textit{isospin-preserving} protons, \textit{total protons}, \textit{isospin-converted} and \textit{produced} protons for both Cascade and MFD modes. The ordering among protons with an exception of \textit{produced} protons starts to diminish above $\sNN\approx$~4 GeV.

\begin{figure*}
\includegraphics[scale=0.38]{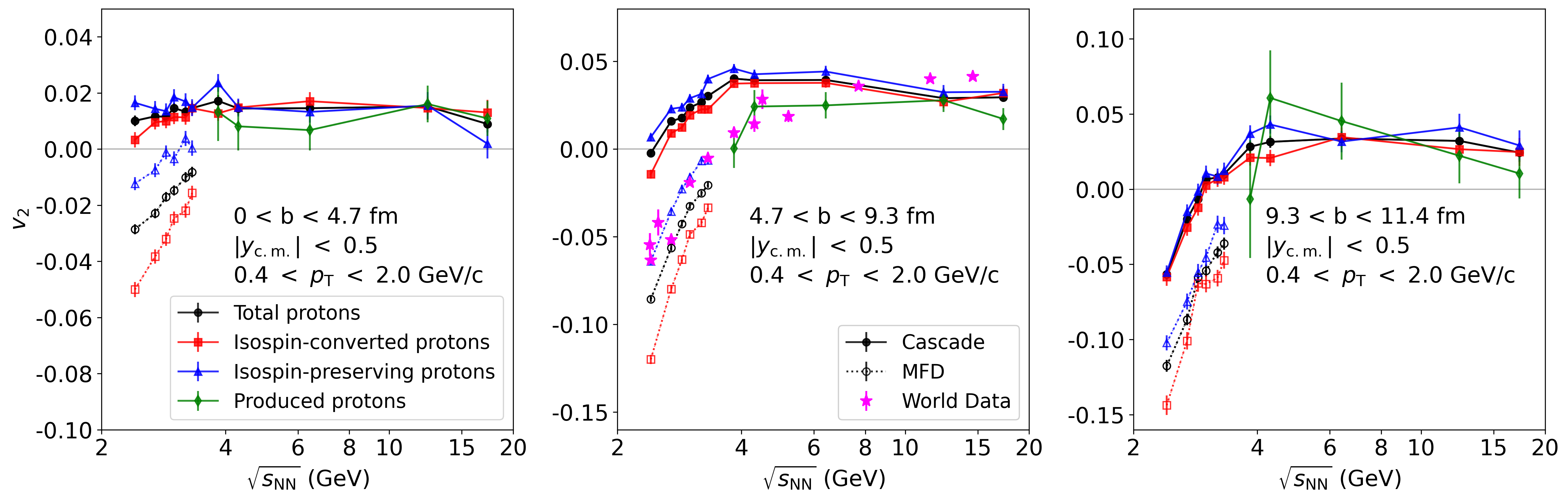}
\caption{Comparison of UrQMD prediction of elliptic flow of \textit{isospin-preserving}, \textit{isospin-converted}, \textit{transported} and \textit{produced} protons as a function of $\sNN$ in different impact parameter collisions in both MFD and Cascade mode. The predictions are compared with experimental measurements in non-central collisions (middle plot).}
\label{f11}
\end{figure*}

The elliptic flow of all categorized protons is also predicted at all energies and impact parameter intervals, as shown in Figure~\ref{f11}. Similar to the slope of directed flow, the predictions are compared with experimental measurements for non-central collisions as shown in the middle plot~\cite{FOPI:2004bfz,E895:2000maf,HADES:2020lob,STAR:2020dav,STAR:2014clz,E895:1999ldn}. The MFD mode reproduces the negative $v_{2}$ at all energies and centralities. In the MFD and Cascade mode overlapping energy range, the MFD predictions for \textit{isospin-converted} protons are consistent with experimental measurements. However, above $\sNN\approx$~3.5 GeV, the data and Cascade predictions agree with each other, except for a brief range around $\sNN\approx$~4 GeV, where the predictions from \textit{produced} protons show reasonable agreement. In central collisions, only MFD predictions show negative $v_{2}$, while Cascade predictions remain positive and flat for the whole energy range. Above $\sNN\approx$~4 GeV, predictions of all categorized protons tends to agree with each other within uncertainties at all energies and impact parameter intervals. A similar ordering is observed in elliptic flow, consistent with origin of differently categorized protons. The \textit{isospin-preserving} protons, originating from initial protons present throughout the whole evolution, exhibit the strongest $v_{2}$ experiencing full pressure gradient. In contrast, \textit{produced} protons, inherit the least flow as they are created predominantly in the later stages of the collisions. The intermediate flow for \textit{isospin-converted} protons reflects their conversion at the later stages of the evolution compared to \textit{isospin-preserving} protons. As a result, total proton $v_{2}$ seems to be diluted in comparison to \textit{isospin-preserving} protons, suggesting implications for experimentally measured total proton flow.

\begin{figure*}
\includegraphics[scale=0.42]{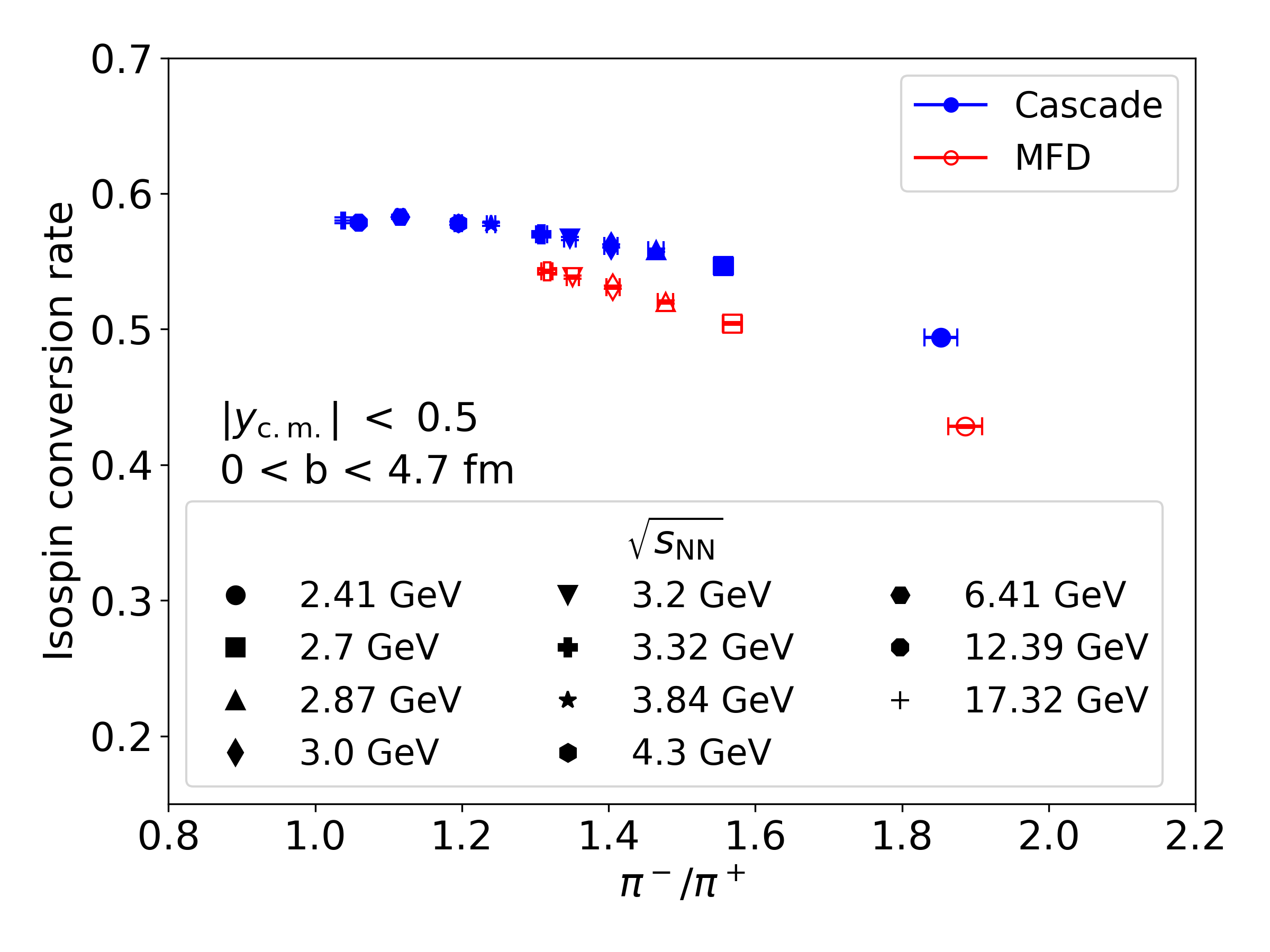}
\includegraphics[scale=0.42]{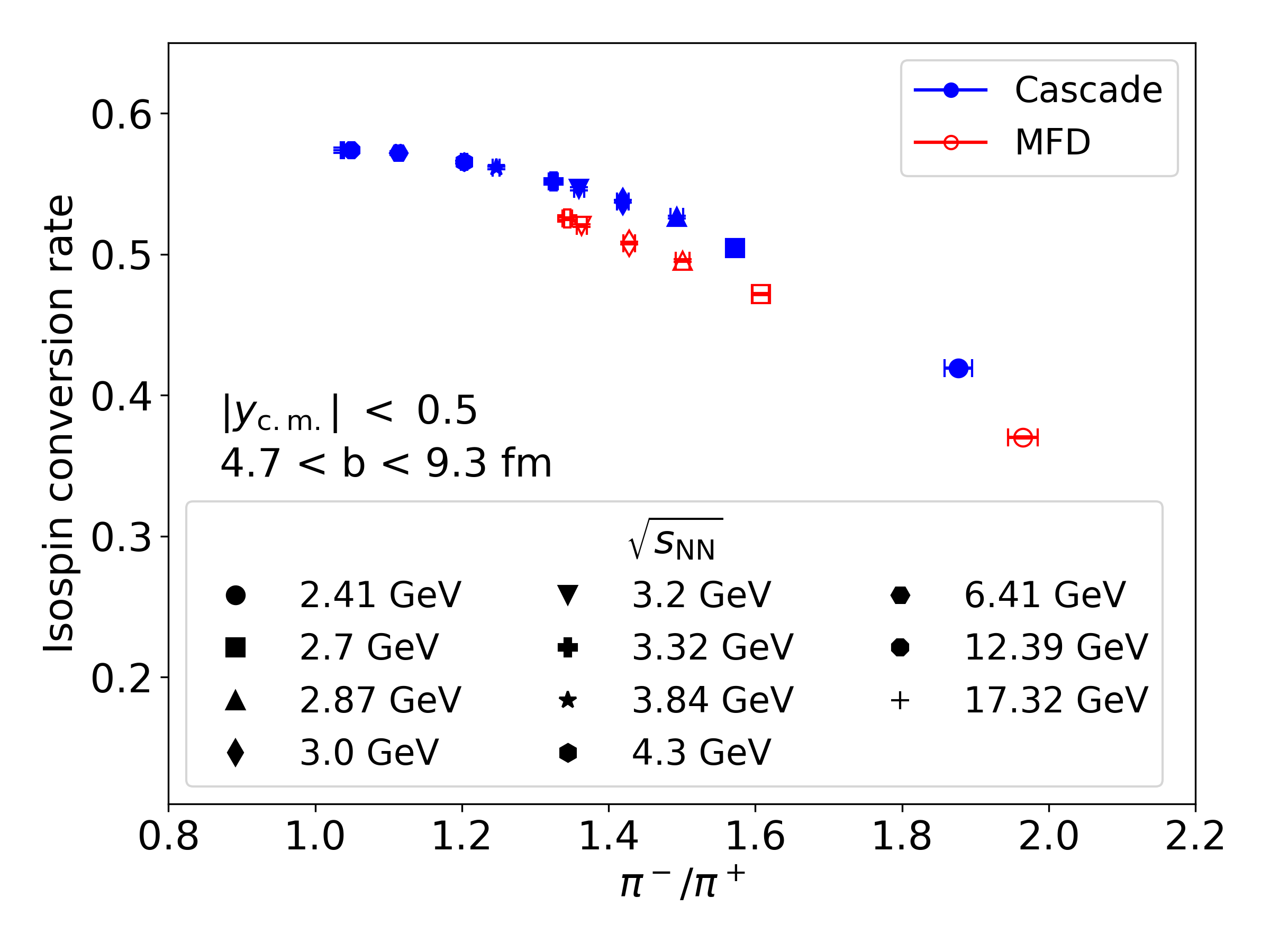}\\
\includegraphics[scale=0.42]{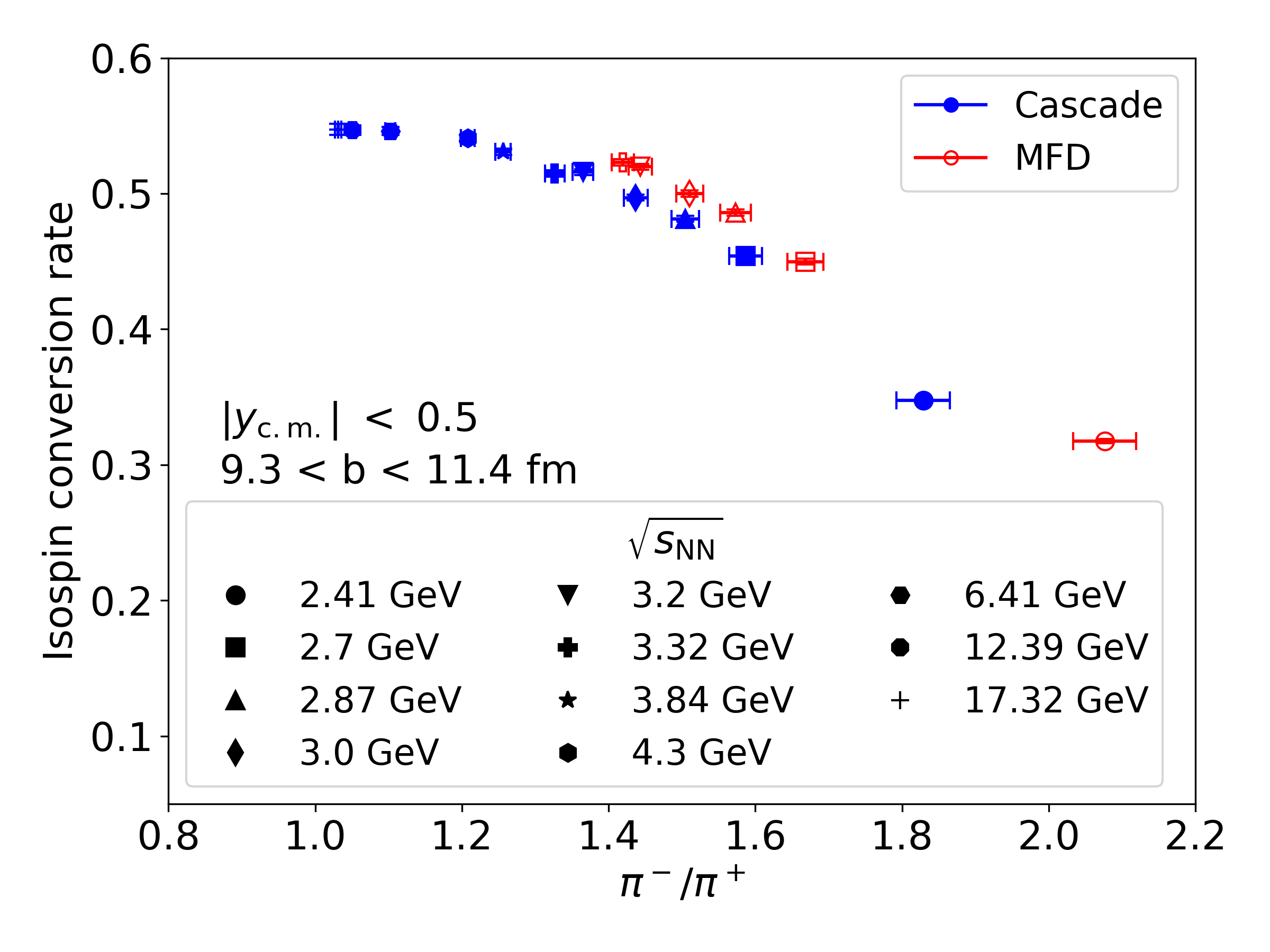}
\includegraphics[scale=0.42]{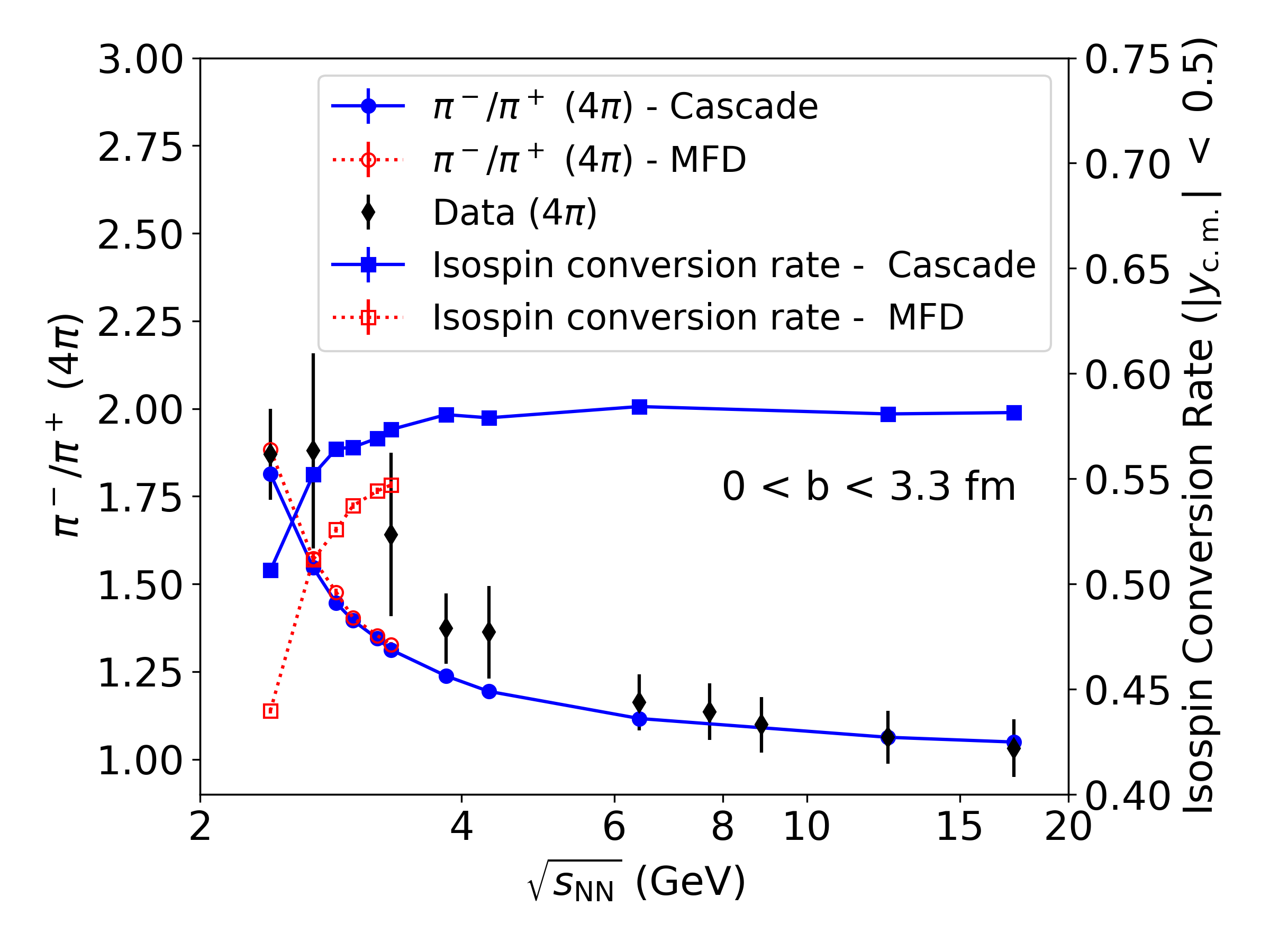}
\caption{Comparison between charged pion ratio ($\pi^{-}/\pi^{+}$) and \textit{isospin conversion} rate in different $\sNN$ and impact parameter intervals (0 $<b<$ 4.7 fm, 4.7 $<b<$ 9.3 fm and 9.3 $<b<$ 11.4 fm) using UrQMD model with MFD and Cascade mode for three impact parameter regions (top panels and bottom left plot). The experimentally measured pion ratios in 4$\pi$ phase space are compared with UrQMD predictions as a function of beam energy along with \textit{isospin conversion} rate at mid-rapidity in central collisions in the bottom right plot.}
\label{f12}
\end{figure*}

Now, the attempt is made to establish a connection between the isospin conversion and experimentally measurable observable. Pion production is known to be sensitive to the isospin dynamics in heavy-ion collisions and charged pion ratio, $\pi^{-}/\pi^{+}$ and is one of most experimentally accessible observables~\cite{Xu:2009fj,HADES:2020ver}. Thus, one compares this ratio with the \textit{isospin conversion} rate at mid-rapidity ($|y_{\rm c.m.}|<$ 0.5) in different impact parameter regions for different collision energies using both modes as shown in the top panel and bottom left plot of Figure~\ref{f12}. From the figure, it is evident that the predicted pion ratio is inversely proportional to \textit{isospin conversion} rate at low energies with ratio slowly approaching unity while \textit{isospin conversion} rate approaching saturation, providing a direct link between pion production and isospin dynamics. The UrQMD predictions for pion ratio are compared with the experimentally measured pion ratios in full phase space~\cite{HADES:2020ver,Klay:2003zf,NA4920,NA4940} in central collisions at AGS and SPS energies, as shown in the bottom right plot of Figure~\ref{f12}. The centrality used for the prediction is comparable to those of AGS and SPS data: the prediction is in 0-5$\%$ (0 $<$ b $<$ 3.3 fm) central collisions, whereas, AGS is in 0-5$\%$ and SPS is in 0-7$\%$ centrality.  The predictions agree with the measured ratios above $\sNN\approx$~6 GeV within experimental uncertainties, with tendency for agreement below this energy. The comparison also reveals two distinct regimes between pion ratio and \textit{isospin conversion} rate separated by $\sNN\approx$ 4 GeV. Above this energy, both observables exhibit a hint of saturation with an increase in energy, suggesting steady isospin conversion through symmetric pion production as pion ratio starts approaching unity. Below this energy, these two observables exhibit opposing behavours, with pion ratio rising to $\approx$~2 at 2.4 GeV, reflecting the neutron-rich nature of Au-Au system, while \textit{isospin conversion} rate decreasing, possibly due to limited pion density.

\begin{figure*}
\includegraphics[scale=0.42]{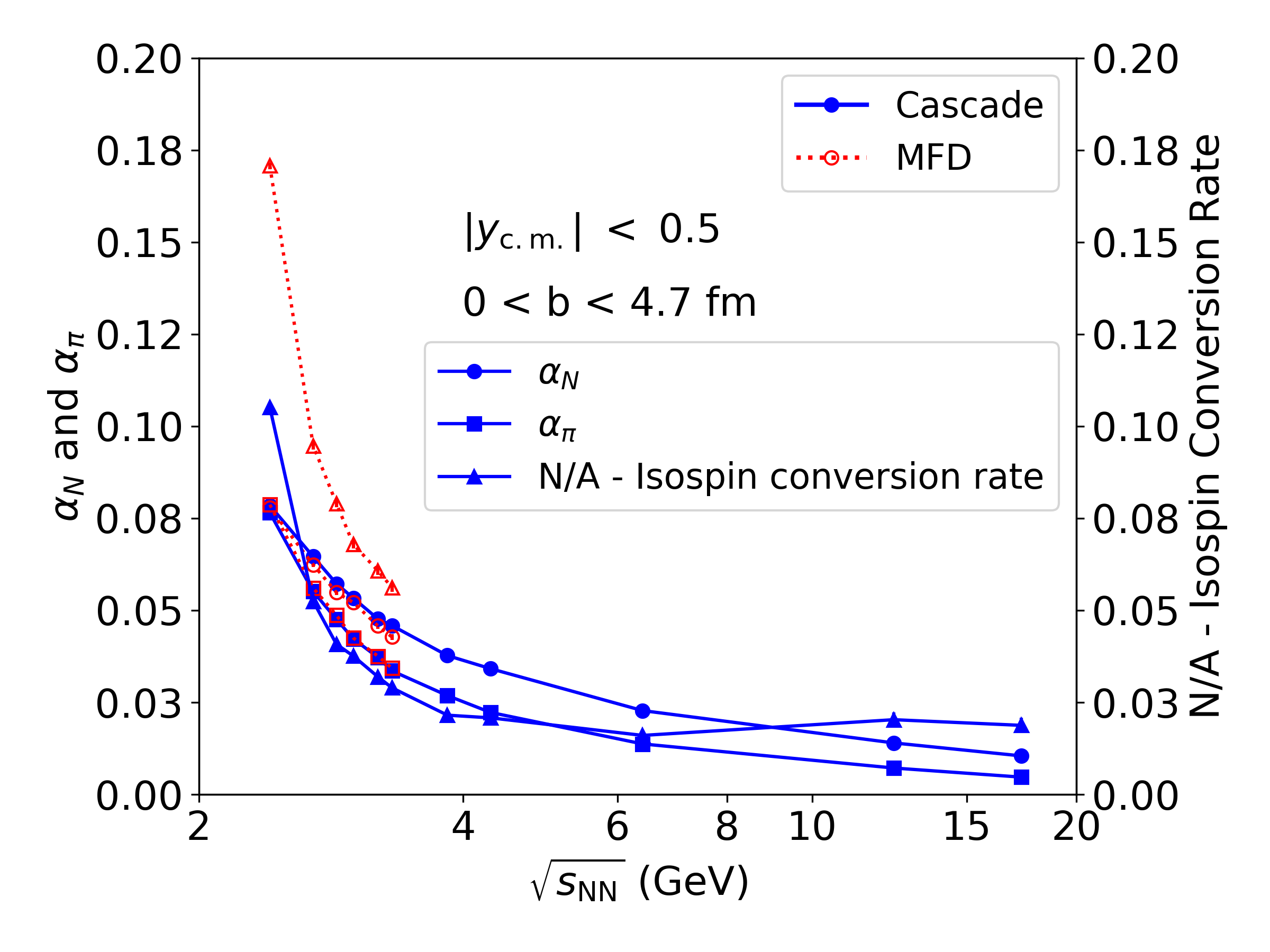}
\includegraphics[scale=0.42]{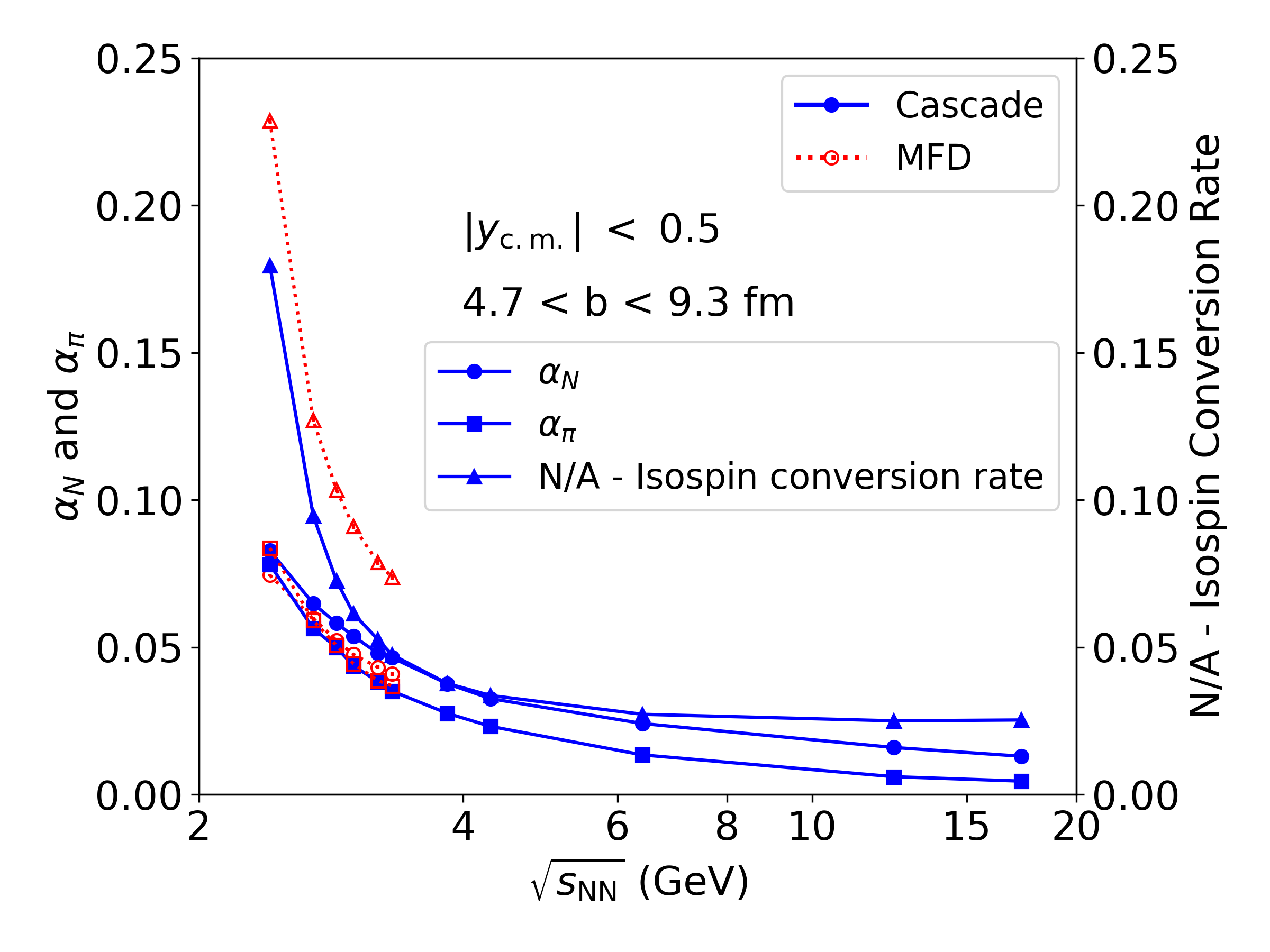}
\includegraphics[scale=0.42]{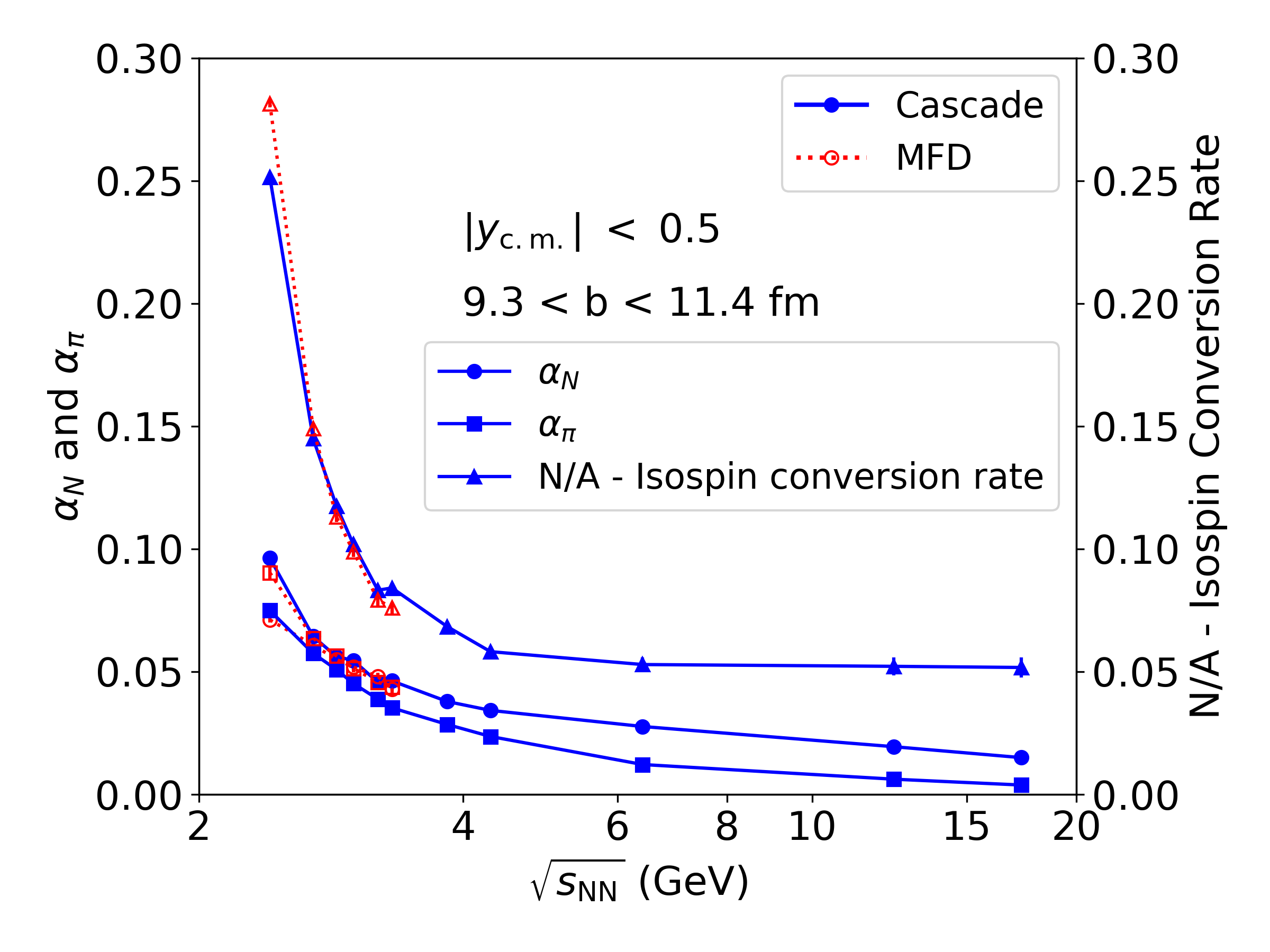}
\caption{Comparison between modified \textit{isospin conversion} rate ($N/A$ - 
\textit{isospin conversion} rate) and $\alpha$ parameters at mid-rapidity ($|y_{\rm c.m.}|<$ 0.5) in different $\sNN$ and impact parameter intervals (0 $<b<$ 4.7 fm, 4.7 $<b<$ 9.3 fm and 9.3 $<b<$ 11.4 fm) using UrQMD model with MFD and Cascade mode for three impact parameter regions.}
\label{f13}
\end{figure*}

At last, one makes a direct comparison of \textit{isospin-conversion} rate with $\alpha$ parameters from KA formalism~\cite{Kitazawa:2012at,Kitazawa:2011wh}. The authors propose an $\alpha$ parameter which enters into the corrections that connects net-proton cumulants to net-baryon cumulants. The eqs. 57 and 63 in~\cite{Kitazawa:2012at}, $\alpha_N$ $=$ $\frac{1}{2}\frac{\langle N_{n}\rangle-\langle N_{p}\rangle}{\langle N_{p}\rangle+\langle N_{n}\rangle}$ and $\alpha_\pi$ $=$ $\frac{1}{2}\frac{\sqrt{\langle N_{\pi^{-}}\rangle/\langle N_{\pi^{+}}\rangle}-1}{\sqrt{\langle N_{\pi^{-}}\rangle/\langle N_{\pi^{+}}\rangle}+1}$ are used and these parameters are estimated in the study. Note that the general form of eq. 63 is used instead of approximated one, $\alpha_\pi$ $\approx$ $\frac{1}{8}$*$\frac{\langle N_{\pi^{-}}\rangle}{\langle N_{\pi^{+}}\rangle}-1$. The comparisons for all three impact parameter regions are shown in Fig.~\ref{f13}. The \textit{isospin conversion} rate is modified as a difference between Neutron to nucleon ratio and the \textit{isospin conversion} rate to match the scale of $\alpha$ parameters. The $\alpha$ parameters do not exhibit the saturation trend which is shown by \textit{isospin conversion} rate. Rather they show gradual decrease as energy increases. In central and mid-central collisions, $\alpha_{N}$ and $\alpha_{\pi}$ show statistically significant disagreement across the entire energy range, with the separation largest at intermediate energies around $\sNN \approx$ 3--6~GeV. Agreement between the two parameters is reached only at $\sNN = 2.4$~GeV. In peripheral collisions, the separation is reduced but remains reasonably similar in the whole energy region, with no agreement at any energy. The modified \textit{isospin conversion} rate deviates systematically at all energies as impact parameter increases indicating incomplete isospin randomization. It is interesting to note that out of the three observables, only the \textit{isospin conversion} rate shows very strong sensitivity to the type of nuclear potential in central and mid-central collisions.

\begin{figure*}
\includegraphics[scale=0.42]{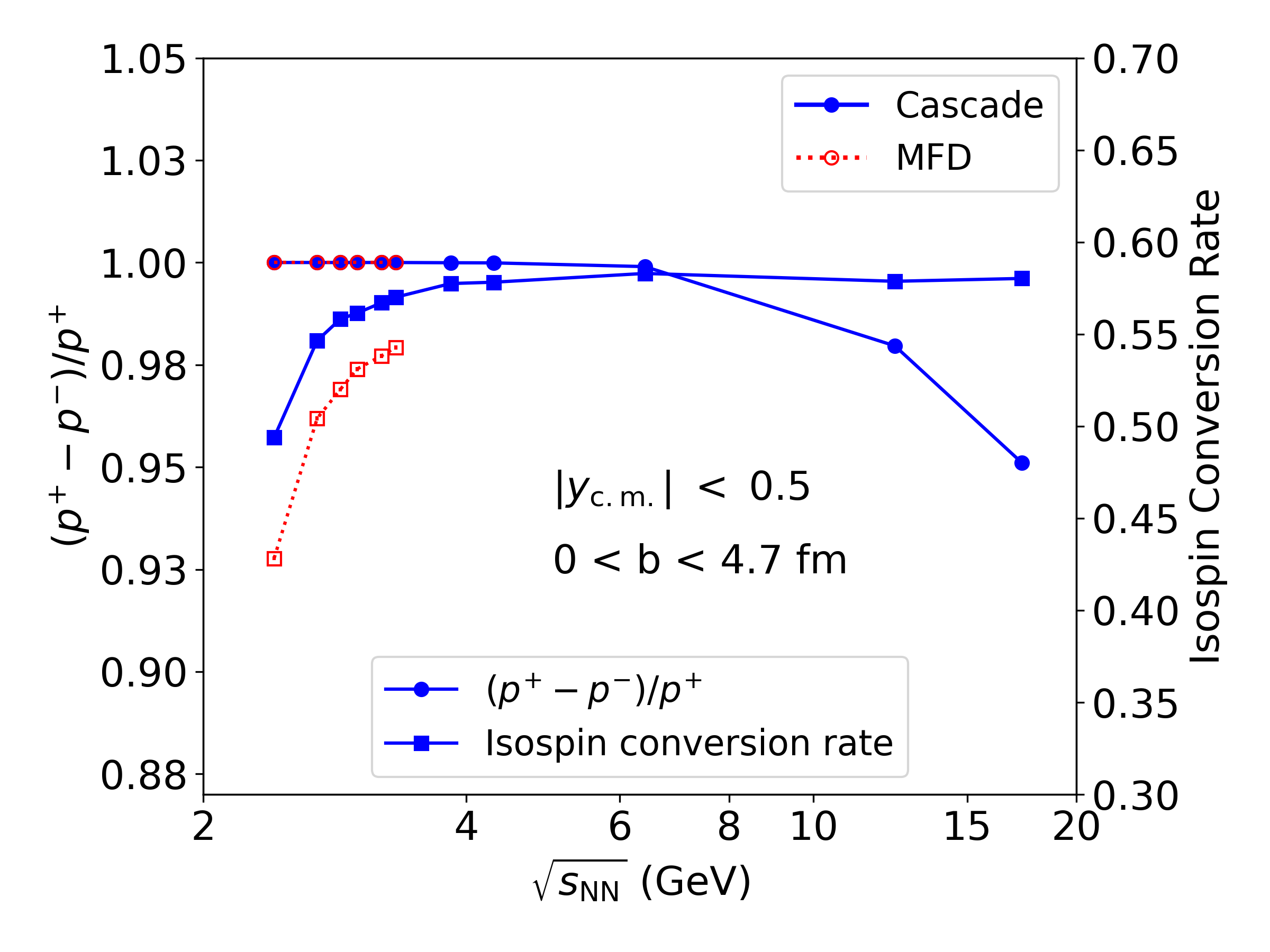}
\includegraphics[scale=0.42]{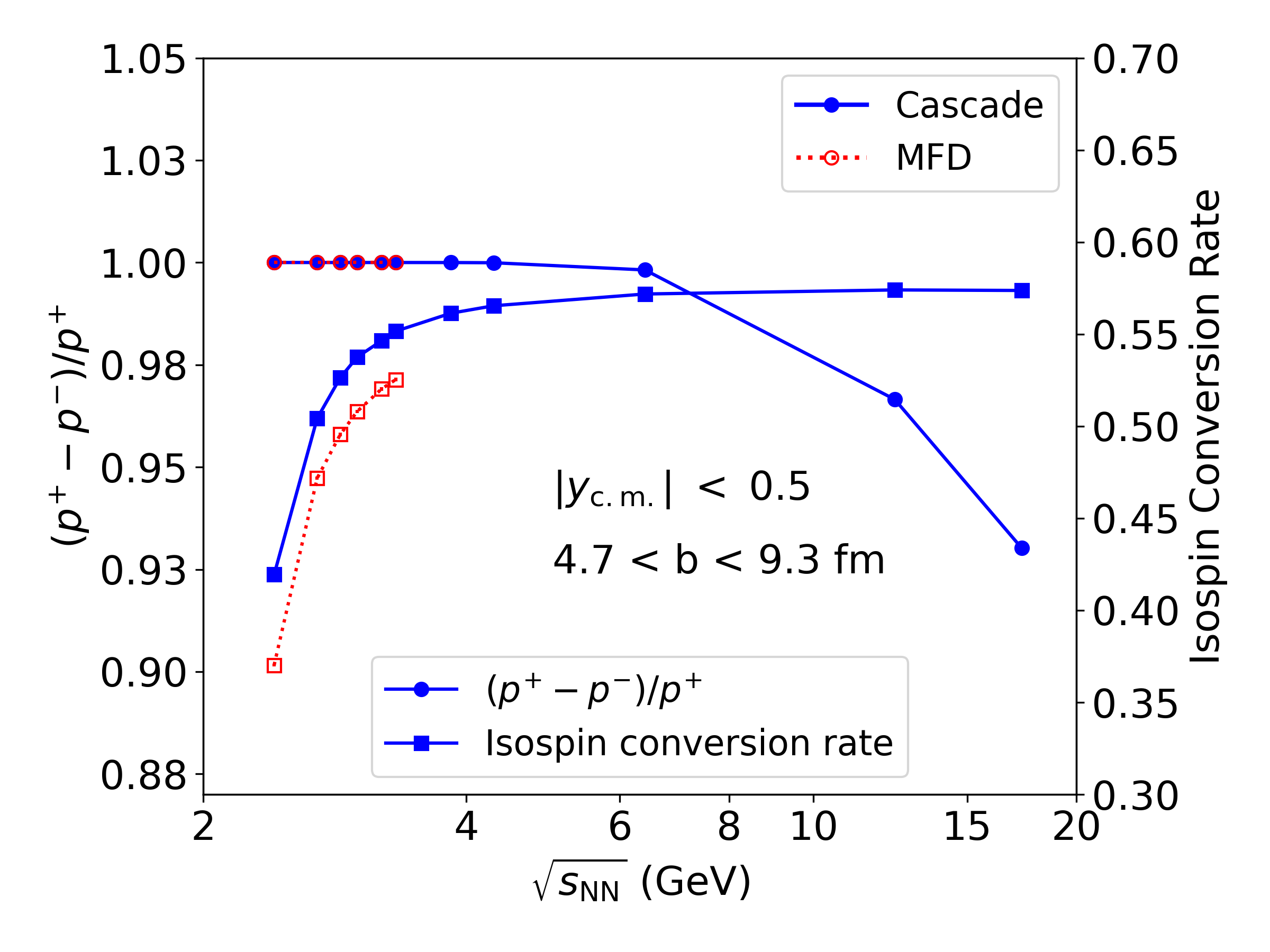}
\includegraphics[scale=0.42]{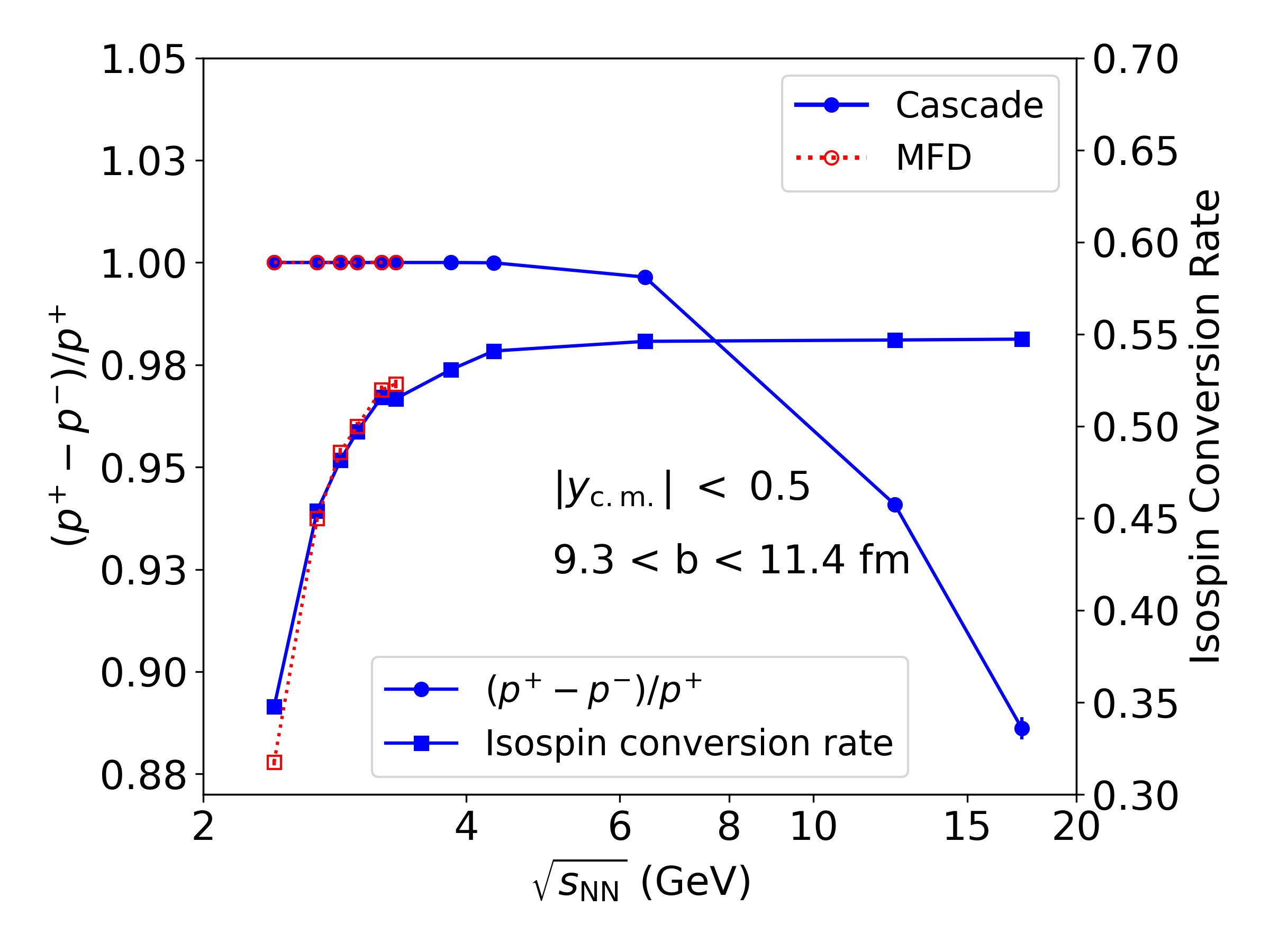}
\caption{Comparison between \textit{isospin conversion} rate and the baryon stopping fraction (net-protons/total protons) at the mid-rapidity ($|y_{\rm c.m.}|<$ 0.5) in different $\sNN$ and impact parameter intervals (0 $<b<$ 4.7 fm, 4.7 $<b<$ 9.3 fm and 9.3 $<b<$ 11.4 fm) using UrQMD model with MFD and Cascade mode for three impact parameter regions.}
\label{f14}
\end{figure*}

In closing, \textit{isospin conversion} rate is compared with the baryon stopping fraction at mid-rapidity ($y_{\rm c.m.}<$~0.5), defined as number of net-protons to the number of protons. The comparison for all three impact parameter regions are shown in Fig.~\ref{f14}. As noted earlier, due to the effect of finite hadronic phase lifetime and limited charge exchange processes, the \textit{isospin conversion} rate saturates at lower values as impact parameter increases, preventing the system from reaching complete isospin randomization. Before making any strong conclusions, a system size study of \textit{isospin conversion} rate is in order. Interestingly, the saturation of the rate coincides with the onset of nuclear transparency between $\sNN \approx$ 4--6~GeV. This demonstrates that isospin randomization and baryon stopping are coupled phenomena in hadronic transport. The degree of isospin randomization achieved is determined by the stopping, saturating when nuclear transparency sets in rather than being solely governed by the charge exchange processes cross section. This transition holds true at all impact parameter regions studied, providing a robust centrality independent feature.

Several limitations of the present study are noted. First, the analysis is performed within a purely hadronic transport framework and does not include a deconfined phase. The results therefore represent a hadronic baseline valid across $\sNN$ = 2.4--17.3~GeV, and provide the reference against which contributions from a QGP phase may be identified at higher energies. Furthermore, the classification of final-state protons into \textit{isospin-preserving} and \textit{isospin-converted} protons relies on simulated interaction history and is not experimentally accessible in real collisions. Therefore, the observable should be understood as a hadronic-transport prediction and can be constrained only indirectly through the experimentally accessible observables reported here, such as, charged pion ratio, the Kitazawa-Asakawa parameters and baryon stopping fraction.  Second, light nuclei production is incorporated in UrQMD~4.0 but not in version~3.4 used in this study. A direct comparison at $\sNN$ = 3.32 and 12.3~GeV shows a difference of $\approx$3$\%$ at the lowest energy and negligible difference at higher energies, confirming that light nuclei production has minimal impact on the results. A detailed study of the isospin conversion rate across the full energy range within UrQMD~4.0, and its effect on light nuclei production, is left for future work. Finally, comparison with other hadronic transport models such as SMASH~\cite{SMASH:2016zqf}, AMPT~\cite{Lin:2004en}, and JAM~\cite{Nara:2019crj} would quantify the model dependence of the findings and is identified as a natural extension of this work.

\section{Conclusion}
In this work, a systematic investigation of microscopic nature of baryon stopping in minimum bias Au+Au collisions over a wide range of collision energies, $\sNN=$ 2.4–17.3 GeV, within the UrQMD transport model is presented. The whole collision history of final-state protons is investigated and the final-state protons are decomposed into distinct categories -- produced protons, \textit{transported} protons, \textit{isospin-preserving} protons and \textit{isospin-converted} protons. Along with Cascade mode across the full energy range, MFD mode of UrQMD has been employed at lower energies. 

In this investigation, it is found that a substantial amount of \textit{transported} protons are originating from initial neutrons, contributions of which are experimentally indistinguishable. This opens up new doors for investigation, as it suggests that conventional net-proton measurements at mid-rapidity also include contributions from \textit{isospin-converted} protons. Therefore, this study provides an opportunity to explore the microscopic nature of \textit{transported} protons and isospin conversion in the said beam energy range and its impact on some of the observables, such as anisotropic flow. 

The yields and fractions of \textit{transported}, \textit{isospin-preserving}, \textit{isospin-converted} and \textit{produced} protons were studied as a function of collision energy for various impact parameter regions at mid-rapidity and found to be dependent on collision energy. Moreover, the estimated \textit{isospin conversion} rate found to be increasing as energy increases and saturating above $\sNN\approx 4$ GeV in all impact parameter regions and was also found to be higher than that of \textit{isospin preservation} rate. The small rate at low energies could be attributed to limited pion density, whereas, saturation above $\sNN\approx 4$ GeV could be due to increase in symmetric pion production. The stopped/transported proton fraction, absolute and relative rapidity loss were studied as a function of beam energy, and it is found that experimental results and predictions for net-protons agree with each other at most energies. The \textit{isospin-converted} protons underwent relatively larger rapidity loss compared to \textit{isospin-preserving} protons. Agreement in the rapidity loss of net-protons and \textit{isospin-converted} protons suggests that the latter dominates the overall rapidity loss in net-protons rather than \textit{isospin-preserving} protons. The average collision time of the isospin conversion, referred to as \textit{flip time}, found to be dependent on the type of conversions, direct or indirect as well as the rapidity region, with larger time is predicted in the forward rapidity region. The investigation suggests dominant direct conversion at low energy with indirect conversion taking over at high energies.

The rapidity spectra of protons in different categories for various energies show interesting behaviour of no forward peaks in the case of \textit{isospin-converted} protons, suggesting more rapidity loss, deflecting the particle to the mid-rapidity. The predictions at all energies overestimate the measured spectra. The shape of the rapidity spectra is quantified in terms of reduced curvature and the predictions found to be reasonably close to extracted measured values at some energies, with deviations at high energies. Reduced curvature shows sensitivity to different categories of protons at low energies, consistent with an origin-based ordering, before becoming indistinguishable above $\sNN\approx$ 4 GeV. 

The slope of directed flow ($dv_{1}/dy$) and elliptic flow ($v_{2}$) of the different proton categories are studied as a function of collision energy in various impact parameter intervals. For the slope of directed flow, Cascade mode predictions mostly overestimate the experimental measurements at intermediate energies with reasonable agreement at the lowest and highest energies. The MFD predictions at 3 GeV are consistent with the data and are consistently higher than Cascade mode across all energies and impact parameter intervals. For $v_{2}$, MFD predictions for \textit{isospin-converted} protons are consistent with experimental data, while Cascade predictions agree with data above $\sNN$ $\approx 6$ GeV. In central collisions, MFD predictions reproduce negative $v_{2}$ while Cascade predictions remain positive across the full energy range, demonstrating significant role of nuclear MFD at low collision energies. A strong origin-based ordering is observed below $\sNN$ $\approx 4$ GeV in both observables with \textit{isospin-preserving} protons show the strongest flow followed by total, \textit{isospin-converted} and \textit{produced} protons.

In order to establish a direct connection of the \textit{isospin conversion} rate with experimental observables, comparisons with the $\pi^{-}/\pi^{+}$ ratio, the $\alpha$ parameters of the Kitazawa-Asakawa formalism~\cite{Kitazawa:2012at,Kitazawa:2011wh}, and the baryon stopping fraction were performed. The \textit{isospin conversion} rate was found to deviate significantly from the KA chemical equilibrium assumption $\alpha_N = \alpha_\pi$ below $\sNN \lesssim$ 10~GeV. Furthermore, its saturation coincides with the onset of nuclear transparency, demonstrating that isospin randomization and baryon stopping are coupled phenomena in hadronic transport. A systematic study of system size dependence could further constrain the isospin dynamics explored in this work.

Finally, the results raise a broader physics question for future investigation: does isospin randomization saturate before, simultaneously with, or after the onset of 
deconfinement? The onset of isospin conversion rate saturation as well as nuclear transparency at $\sNN \approx$ 4--6~GeV coincides with the energy region where signatures of deconfinement onset have been discussed in the literature~\cite{Ivanov:2010cu,Ivanov:2015vna}. The hadronic transport study performed here establishes a baseline for such a question, and motivates further investigations incorporating a deconfined phase in this energy region.

\section{Acknowledgements}

The author is deeply grateful to Ankhi Roy and Ramin Barak for their critical reading of the manuscript. The author thanks Ramin Barak for many fruitful discussions.


\begin{thebibliography}{9} 

\bibitem{Florkowski:2014yza} 
  W.~Florkowski,
  %``Basic phenomenology for relativistic heavy-ion collisions,''
  Acta Phys.\ Polon.\ B {\bf 45}, no. 12, 2329 (2014).

\bibitem{UWHeinz} U. W. Heinz, Concepts of heavy-ion physics, \textit{2nd CERN-
CLAF School of High Energy Physics} (2004), pp. 165–238;
arXiv:hep-ph/0407360.

\bibitem{BraunMunzinger:2008tz} 
  P.~Braun-Munzinger and J.~Wambach,
  %``The Phase Diagram of Strongly-Interacting Matter,''
  Rev.\ Mod.\ Phys.\  {\bf 81}, 1031 (2009).
 % doi:10.1103/RevModPhys.81.1031
 % [arXiv:0801.4256 [hep-ph]].

\bibitem{rhic1} J. Adams {\it et. al.} [STAR Collaboration], Nucl. Phys. A {\bf 757} (2005), 102. 

\bibitem{rhic2} K. Adcox {\it et. al.} [PHENIX Collaboration], Nucl. Phys. A {\bf 757} (2005), 184.

\bibitem{lhc1} K. Aamodt {\it et. al.} [ALICE Collaboration], Phys. Rev. Lett. {\bf 107} (2011), 032301.

\bibitem{lhc2} G. Aad {\it et. al.} [ATLAS Collaboration], Phys. Rev. C {\bf 86}
 (2012), 014907.

\bibitem{lhc3} S. Chatrchyan {\it et. al.} [CMS Collaboration], Phys. Rev. C {\bf 89} (2014), 044906.

\bibitem{sQGP} E. Shuryak, Rev. Mod. Phys. {\bf 89} (2017) 035001;
U.~Heinz, C.~Shen and H.~Song,
  %``The viscosity of quark-gluon plasma at RHIC and the LHC,''
  AIP Conf.\ Proc.\  {\bf 1441} (2012) 766.
%  doi:10.1063/1.3700674
%  [arXiv:1108.5323 [nucl-th]].

%\cite{CBM:2016kpk}
\bibitem{CBM:2016kpk}
T.~Ablyazimov \textit{et al.} [CBM],
%``Challenges in QCD matter physics --The scientific programme of the Compressed Baryonic Matter experiment at FAIR,''
Eur. Phys. J. A \textbf{53} (2017) no.3, 60
doi:10.1140/epja/i2017-12248-y
[arXiv:1607.01487 [nucl-ex]].
%401 citations counted in INSPIRE as of 08 Dec 2025
  
\bibitem{nica} 
  V.~Toneev,
  %``The NICA/MPD project at JINR (Dubna),''
  PoS CPOD {\bf 07}, 057 (2007).
 % [arXiv:0709.1459 [nucl-ex]].
 
%\cite{MPD:2022qhn}
\bibitem{MPD:2022qhn}
V.~Abgaryan \textit{et al.} [MPD],
%``Status and initial physics performance studies of the MPD experiment at NICA,''
Eur. Phys. J. A \textbf{58} (2022) no.7, 140
doi:10.1140/epja/s10050-022-00750-6
[arXiv:2202.08970 [physics.ins-det]].
%91 citations counted in INSPIRE as of 08 Dec 2025

%\cite{MPD:2025jzd}
\bibitem{MPD:2025jzd}
R.~Abdulin \textit{et al.} [MPD],
%``MPD physics performance studies in Bi+Bi collisions at {\ensuremath{\sqrt{}}}sNN = 9.2 GeV,''
Rev. Mex. Fis. \textbf{71} (2025) no.4, 041201
doi:10.31349/RevMexFis.71.041201
[arXiv:2503.21117 [nucl-ex]].
%4 citations counted in INSPIRE as of 08 Dec 2025 


%\cite{Mohs:2019xvz}
\bibitem{Mohs:2019xvz}
J.~Mohs, S.~Ryu and H.~Elfner,
%``Can Baryon Stopping Be Understood within a Hadronic Transport Approach,''
MDPI Proc. \textbf{10} (2019) no.1, 2
doi:10.3390/proceedings2019010002
%0 citations counted in INSPIRE as of 20 Mar 2026

%\cite{Ivanov:2016xev}
\bibitem{Ivanov:2016xev}
Y.~B.~Ivanov, D.~Blaschke,
%``Baryon stopping in heavy-ion collisions at E$_{lab}$ = 2A-200A GeV,''
Eur. Phys. J. A 52 (2016) no.8 237.
%doi:10.1140/epja/i2016-16237-4
%4 citations counted in INSPIRE as of 01 May 2021


\bibitem{NA4920}
 C. Alt {\it et. al.} NA49 Collaboration, Phys.Rev. C77 (2008) 024903, 2008 .

\bibitem{NA4940}
 S. V. Afanasiev {\it et. al.} NA49 Collaboration, Phys.Rev. C66 (2002) 054902, 2002 .

\bibitem{NA49158}
C. Alt {\it et. al.} NA49 Collaboration, Phys.Rev. C68 (2003) 034903.

\bibitem{Klay:2003zf} 
  J.~L.~Klay {\it et al.} [E-0895 Collaboration],
  %``Charged pion production in 2 to 8 agev central au+au collisions,''
  Phys. Rev. C {\bf 68}, 054905 (2003)

\bibitem{Klay:2001tf} 
  J.~L.~Klay {\it et al.} [E895 Collaboration],
  %``Longitudinal flow from 2-A-GeV to 8-A-GeV Au+Au collisions at the Brookhaven AGS,''
  Phys.\ Rev.\ Lett.\  {\bf 88}, 102301 (2002)
  %doi:10.1103/PhysRevLett.88.102301 [nucl-ex/0111006].
  %%CITATION = doi:10.1103/PhysRevLett.88.102301;%%
  %75 citations counted in INSPIRE as of 26 Jan 2018

%\cite{Back:2000ru}
\bibitem{Back:2000ru}
B.~B.~Back, et al., [E917],
%``Baryon rapidity loss in relativistic Au+Au collisions,''
Phys. Rev. Lett. 86 (2001) 1970-1973.
%doi:10.1103/PhysRevLett.86.1970
%[arXiv:nucl-ex/0003007 [nucl-ex]].
%120 citations counted in INSPIRE as of 01 May 2021

%\cite{BRAHMS:2009wlg}
\bibitem{BRAHMS:2009wlg}
I.~C.~Arsene \textit{et al.} [BRAHMS],
%``Nuclear stopping and rapidity loss in Au+Au collisions at s(NN)**(1/2) = 62.4-GeV,''
Phys. Lett. B \textbf{677} (2009), 267-271
doi:10.1016/j.physletb.2009.05.049
[arXiv:0901.0872 [nucl-ex]].
%105 citations counted in INSPIRE as of 20 Mar 2026



\bibitem{bass1998}
S.~A.~Bass \textit{et al.}, 
``Microscopic models for ultrarelativistic heavy ion collisions,'' 
Prog.\ Part.\ Nucl.\ Phys.\ {\bf 41}, 225–370 (1998).

\bibitem{bleicher1999}
M.~Bleicher \textit{et al.}, 
``Relativistic hadron-hadron collisions in the ultrarelativistic quantum molecular dynamics model,'' 
J.\ Phys.\ G {\bf 25}, 1859–1896 (1999).

%\cite{SMASH:2016zqf}
\bibitem{SMASH:2016zqf}
J.~Weil \textit{et al.} [SMASH],
%``Particle production and equilibrium properties within a new hadron transport approach for heavy-ion collisions,''
Phys. Rev. C \textbf{94} (2016) no.5, 054905
doi:10.1103/PhysRevC.94.054905
[arXiv:1606.06642 [nucl-th]].
%414 citations counted in INSPIRE as of 20 Mar 2026



%\cite{Weber:2002qb}
\bibitem{Weber:2002qb}
H.~Weber, E.~L.~Bratkovskaya and H.~Stoecker,
%``Baryon stopping and strange baryon and anti-baryon production at ultrarelativistic energies,''
Phys. Rev. C \textbf{66} (2002), 054903
doi:10.1103/PhysRevC.66.054903
%19 citations counted in INSPIRE as of 20 Mar 2026



%\cite{Mohs:2019iee}
\bibitem{Mohs:2019iee}
J.~Mohs \textit{et al.} [SMASH],
%``Particle Production via Strings and Baryon Stopping within a Hadronic Transport Approach,''
J. Phys. G \textbf{47} (2020) no.6, 065101
doi:10.1088/1361-6471/ab7bd1
[arXiv:1909.05586 [nucl-th]].
%69 citations counted in INSPIRE as of 22 Mar 2026

%\cite{Ivanov:2010cu}
\bibitem{Ivanov:2010cu}
Y.~B.~Ivanov,
%``Baryon Stopping in Heavy-Ion Collisions at $E_{lab}$ = 2-160 GeV/nucleon,''
Phys. Lett. B 690 (2010) 358-362.
%doi:10.1016/j.physletb.2010.05.051
%[arXiv:1001.0670 [nucl-th]].
%18 citations counted in INSPIRE as of 01 May 2021

%\cite{Ivanov:2012bh}
\bibitem{Ivanov:2012bh}
Y.~B.~Ivanov,
%``Baryon Stopping as a Probe of Deconfinement Onset in Relativistic Heavy-Ion Collisions,''
Phys. Lett. B 721 (2013) 123-130.
%doi:10.1016/j.physletb.2013.02.038
%[arXiv:1211.2579 [hep-ph]].
%27 citations counted in INSPIRE as of 01 May 2021

%\cite{Ivanov:2013mxa}
\bibitem{Ivanov:2013mxa}
Y.~B.~Ivanov,
%``Elliptic Flow of Protons and Antiprotons in Au+Au Collisions at $\sNN$ = 7.7--62.4 GeV within Alternative Scenarios of Three-Fluid Dynamics,''
Phys. Lett. B 723 (2013) 475-480.
%doi:10.1016/j.physletb.2013.05.053
%[arXiv:1304.2307 [nucl-th]].
%21 citations counted in INSPIRE as of 01 May 2021


%\cite{Ivanov:2015vna}
\bibitem{Ivanov:2015vna}
Y.~B.~Ivanov, D.~Blaschke,
%``Robustness of the Baryon-Stopping Signal for the Onset of Deconfinement in Relativistic Heavy-Ion Collisions,''
Phys. Rev. C 92 (2015) no.2 024916.
%doi:10.1103/PhysRevC.92.024916
%[arXiv:1504.03992 [nucl-th]].
%24 citations counted in INSPIRE as of 01 May 2021

%\cite{Thakur:2016znw}
\bibitem{Thakur:2016znw}
D.~Thakur, S.~Jakhar, P.~Garg and R.~Sahoo,
%``Estimation of Stopped Protons at Energies Relevant for a Beam Energy Scan at the BNL Relativistic Heavy-Ion Collider,''
Phys. Rev. C \textbf{95} (2017) no.4, 044903
doi:10.1103/PhysRevC.95.044903
[arXiv:1611.05078 [nucl-ex]].
%10 citations counted in INSPIRE as of 20 Mar 2026


 
%\cite{Zhong:2010zz}
\bibitem{Zhong:2010zz}
Y.~Zhong and S.~Q.~Feng,
%``Rapidity distributions of net protons from AGS to LHC energy regions,''
Chin. Phys. C \textbf{34} (2010), 1085-1089
doi:10.1088/1674-1137/34/8/009
%0 citations counted in INSPIRE as of 20 Mar 2026 




%\cite{Kuiper:2006si}
\bibitem{Kuiper:2006si}
R.~Kuiper and G.~Wolschin,
%``From RHIC to LHC: A Relativistic diffusion approach,''
Annalen Phys. \textbf{16} (2007), 67-77
doi:10.1002/andp.200610225
[arXiv:hep-ph/0610195 [hep-ph]].
%10 citations counted in INSPIRE as of 20 Mar 2026

%\cite{Kitazawa:2012at}
\bibitem{Kitazawa:2012at}
M.~Kitazawa and M.~Asakawa,
%``Relation between baryon number fluctuations and experimentally observed proton number fluctuations in relativistic heavy ion collisions,''
Phys. Rev. C \textbf{86} (2012), 024904
[erratum: Phys. Rev. C \textbf{86} (2012), 069902]
doi:10.1103/PhysRevC.86.024904
[arXiv:1205.3292 [nucl-th]].
%205 citations counted in INSPIRE as of 01 Apr 2026

%\cite{Kitazawa:2011wh}
\bibitem{Kitazawa:2011wh}
M.~Kitazawa and M.~Asakawa,
%``Revealing baryon number fluctuations from proton number fluctuations in relativistic heavy ion collisions,''
Phys. Rev. C \textbf{85} (2012), 021901
doi:10.1103/PhysRevC.85.021901
[arXiv:1107.2755 [nucl-th]].
%151 citations counted in INSPIRE as of 01 Apr 2026

\bibitem{manual}
UrQMD manual, \url{https://itp.uni-frankfurt.de/~bleicher/UrQMD_user_manual.pdf}

%\cite{HADES:2017def}
\bibitem{HADES:2017def}
J.~Adamczewski-Musch \textit{et al.} [HADES],
%``Centrality determination of Au + Au collisions at 1.23A GeV with HADES,''
Eur. Phys. J. A \textbf{54} (2018) no.5, 85
doi:10.1140/epja/i2018-12513-7
[arXiv:1712.07993 [nucl-ex]].
%75 citations counted in INSPIRE as of 06 Apr 2026




%\cite{FOPI:2004bfz}
\bibitem{FOPI:2004bfz}
A.~Andronic \textit{et al.} [FOPI],
%``Excitation function of elliptic flow in Au+Au collisions and the nuclear matter equation of state,''
Phys. Lett. B \textbf{612} (2005), 173-180
doi:10.1016/j.physletb.2005.02.060
[arXiv:nucl-ex/0411024 [nucl-ex]].
%206 citations counted in INSPIRE as of 24 Apr 2026

%\cite{E895:2000maf}
\bibitem{E895:2000maf}
H.~Liu \textit{et al.} [E895],
%``Sideward flow in Au + Au collisions between 2-A-GeV and 8-A-GeV,''
Phys. Rev. Lett. \textbf{84} (2000), 5488-5492
doi:10.1103/PhysRevLett.84.5488
[arXiv:nucl-ex/0005005 [nucl-ex]].
%136 citations counted in INSPIRE as of 24 Apr 2026

%\cite{HADES:2020lob}
\bibitem{HADES:2020lob}
J.~Adamczewski-Musch \textit{et al.} [HADES],
%``Directed, Elliptic, and Higher Order Flow Harmonics of Protons, Deuterons, and Tritons in $\mathrm{Au}+\mathrm{Au}$ Collisions at $\sqrt{{s}_{NN}}=2.4\text{ }\text{ }\mathrm{GeV}$,''
Phys. Rev. Lett. \textbf{125} (2020), 262301
doi:10.1103/PhysRevLett.125.262301
[arXiv:2005.12217 [nucl-ex]].
%102 citations counted in INSPIRE as of 24 Apr 2026

%\cite{STAR:2020dav}
\bibitem{STAR:2020dav}
J.~Adam \textit{et al.} [STAR],
%``Flow and interferometry results from Au+Au collisions at $\sqrt{s_{NN}} = 4.5$ GeV,''
Phys. Rev. C \textbf{103} (2021) no.3, 034908
doi:10.1103/PhysRevC.103.034908
[arXiv:2007.14005 [nucl-ex]].
%108 citations counted in INSPIRE as of 24 Apr 2026

%\cite{STAR:2014clz}
\bibitem{STAR:2014clz}
L.~Adamczyk \textit{et al.} [STAR],
%``Beam-Energy Dependence of the Directed Flow of Protons, Antiprotons, and Pions in Au+Au Collisions,''
Phys. Rev. Lett. \textbf{112} (2014) no.16, 162301
doi:10.1103/PhysRevLett.112.162301
[arXiv:1401.3043 [nucl-ex]].
%322 citations counted in INSPIRE as of 24 Apr 2026

%\cite{E895:1999ldn}
\bibitem{E895:1999ldn}
C.~Pinkenburg \textit{et al.} [E895],
%``Elliptic flow: Transition from out-of-plane to in-plane emission in Au + Au collisions,''
Phys. Rev. Lett. \textbf{83} (1999), 1295-1298
doi:10.1103/PhysRevLett.83.1295
[arXiv:nucl-ex/9903010 [nucl-ex]].
%266 citations counted in INSPIRE as of 24 Apr 2026


%\cite{HADES:2020ver}
\bibitem{HADES:2020ver}
J.~Adamczewski-Musch \textit{et al.} [HADES],
%``Charged-pion production in $\mathbf {Au+Au}$ collisions at $\sqrt{\mathbf {s}_{\mathbf {NN}}} = 2.4~{\mathbf {GeV}}$: HADES Collaboration,''
Eur. Phys. J. A \textbf{56} (2020) no.10, 259
doi:10.1140/epja/s10050-020-00237-2
[arXiv:2005.08774 [nucl-ex]].
%47 citations counted in INSPIRE as of 07 Apr 2026


%\cite{Xu:2009fj}
\bibitem{Xu:2009fj}
J.~Xu, C.~M.~Ko and Y.~Oh,
%``Isospin-dependent pion in-medium effects on charged pion ratio in heavy ion collisions,''
Phys. Rev. C \textbf{81} (2010), 024910
doi:10.1103/PhysRevC.81.024910
[arXiv:0906.1602 [nucl-th]].
%42 citations counted in INSPIRE as of 07 Apr 2026

%\cite{Lin:2004en}
\bibitem{Lin:2004en}
Z.~W.~Lin, C.~M.~Ko, B.~A.~Li, B.~Zhang and S.~Pal,
%``A Multi-phase transport model for relativistic heavy ion collisions,''
Phys. Rev. C \textbf{72} (2005), 064901
%doi:10.1103/PhysRevC.72.064901
%[arXiv:nucl-th/0411110 [nucl-th]].
%1660 citations counted in INSPIRE as of 26 Jun 2026

%\cite{Nara:2019crj}
\bibitem{Nara:2019crj}
Y.~Nara,
%``JAM: an event generator for high energy nuclear collisions,''
EPJ Web Conf. \textbf{208} (2019), 11004.
%doi:10.1051/epjconf/201920811004
%25 citations counted in INSPIRE as of 26 Jun 2026


\end{thebibliography}
\end{document}